\documentclass{article}

\usepackage{natbib}

\renewcommand{\cite}[2][]{\citep[#1]{#2}}
\renewcommand{\citetext}[2][]{\citet[#1]{#2}}
\usepackage[colorlinks,linkcolor=blue,citecolor=blue,urlcolor=blue]{hyperref}
\usepackage[all]{hypcap}    %for going to the top of an image when a figure reference is clicked

\usepackage[makeroom]{cancel}
\usepackage{datetime}
\usepackage[colorlinks,linkcolor=blue,citecolor=blue,urlcolor=blue]{hyperref}
\usepackage{graphicx}
\usepackage{ushort}
\usepackage{amsmath,amssymb}
\usepackage{setspace}

\usepackage[utf8]{inputenc}

\usepackage[LGR,T1]{fontenc}
\usepackage{amsmath,etoolbox}
\usepackage{accents}

\newcommand{\eq}[1]{Eq. \eqref{#1}}
\newcommand{\eqs}[1]{Eqs. \eqref{#1}}
\newcommand{\fig}[1]{Fig. \ref{#1}}

\def\clap#1{\hbox to 0pt{\hss#1\hss}}

\newcommand{\Reals}{\mathbb{R}}
\newcommand{\trsp}{^{\!\mathsf{T}}}
\newcommand{\mtrsp}{^{-\!\mathsf{T}}}

\newcommand{\lfrac}[2]{#1/#2}

\newcommand{\pt}[1]{\boldsymbol{#1}}
\newcommand{\vc}[1]{\pt{\vecteur{#1}}}

\newcommand{\ts}[1]{\boldsymbol{\mathbf{#1}}}

\DeclareMathOperator{\tr}{tr}
\newcommand{\grad}{\vc{\nabla}}
\newcommand{\gradx}{\vc{\nabla}_{\!\pt{x}}}
\newcommand{\gradr}{\vc{\nabla}_{\!\pt{r}}}
\newcommand{\gradrbar}{\vc{\nabla}_{\!\bar{\pt{r}}}}

\renewcommand{\div}{\grad\cdot}

\NewDocumentCommand{\ucm}{m}{\accentset{\triangledown}{\ts{#1}}}

\newcommand{\ded}[2]{\frac{\partial #1}{\partial #2}}
\newcommand{\ddt}[1]{\frac{\mathrm{d} #1}{\mathrm{d} t}}
\newcommand{\dedt}[1]{\frac{\partial #1}{\partial t}}
\newcommand{\lddt}[1]{{\mathrm{d} #1}/{\mathrm{d} t}}
\newcommand{\intd}[4]{\int_{#1\!\!\!}^{#2} #3 \:\text{d}#4}

\ushortCreate()[-.6\ht0]{ushort}
\renewcommand{\pt}[1]{{\boldsymbol{{#1}}}}
\renewcommand{\vc}[1]{\protect\underaccent{\sim}{\boldsymbol{{#1}}}}
\renewcommand{\ts}[1]{\protect\underaccent{\hspace{.1em}\approx}{\boldsymbol{\mathbf{#1}}}}
\newcommand{\tsit}[1]{\protect\underaccent{\approx}{\boldsymbol{{#1}}}}
\renewcommand{\fig}[1]{Fig.~\ref{#1}}
\renewcommand{\eq}[1]{Eq.~(\ref{#1})}
\newcommand{\refsec}[1]{Section~\ref{#1}}
\newcommand{\refapp}[1]{Appendix~\ref{#1}}
\DeclareMathOperator{\dev}{\mathbf{dev}}
\DeclareMathOperator{\diag}{\mathbf{diag}}
\DeclareMathOperator{\Prob}{{Prob}}
\newcommand{\hide}[1]{}
\newcommand{\JEadds}[1]{#1}

\newtheorem{hyp}{Hypothesis}

\newcommand{\tsigma}{\ts{\sigma}}

\newcommand{\dts}[1]{\dot{\ts{#1}}\vphantom{\ts{#1}}}

\newcommand{\tsD}{\tsit{{d}}}
\newcommand{\tsL}{\tsit{{\ell}}}

\newcommand{\Deltamu}{\Delta\mu}

\newcommand{\tsTau}{\ts{\tau}}
\newcommand{\tA}{\ts{A}}
\newcommand{\tsiga}{\ts{\sigma}_\mathrm{a}}
\newcommand{\tAa}{\ts{A}_\mathrm{a}}
\newcommand{\tEa}{\ts{E}_\mathrm{a}}
\newcommand{\Ja}{\vc{j}_\mathrm{\!a}}
\newcommand{\rangleb}{\rangle_\mathrm{b}}
\newcommand{\ku}{k_\mathrm{u}}
\newcommand{\kb}{k_\mathrm{b}}

\newcommand{\ks}{k_\mathrm{s}}
\newcommand{\ka}{k_\mathrm{a}}
\newcommand{\va}{v_\mathrm{a}}
\newcommand{\ella}{\ell_\mathrm{a}}
\newcommand{\au}{a_\mathrm{u}}

\newcommand{\Ru}{\vc{R}_\mathrm{u}}
\newcommand{\Rb}{\vc{R}_\mathrm{b}}
\newcommand{\dRu}{\dot{\vc{R}}_\mathrm{u}}
\newcommand{\dRb}{\dot{\vc{R}}_\mathrm{b}}
\newcommand{\psiu}{\psi_\mathrm{u}}
\newcommand{\psib}{\psi_\mathrm{b}}
\newcommand{\Be}{\ts{B}_{\mathrm{e}}}
\newcommand{\Fe}{\ts{F}_{\!\mathrm{e}}}

\newcommand{\Fva}{\ts{F}_{\!\mathrm{va}}}
\newcommand{\Cva}{\ts{C}_{\!\mathrm{va}}}

\newcommand{\xic}{\xi_\mathrm{a}}
\newcommand{\dxic}{\dot{\xi}_\mathrm{a}}
\begin{document}

\title{Mechanics and thermodynamics of contractile {entropic} biopolymer networks
}

%\titlerunning{Short form of title}        % if too long for running head

\author{Antoine Jallon \and Pierre Recho \and Jocelyn \'Etienne
}

\date{\small
              Univ.\ Grenoble Alpes--CNRS, LIPHY \\
              \texttt{jocelyn.etienne@univ-grenoble-alpes.fr}           
}

\maketitle

\begin{abstract}
Contractile biopolymer networks,
such as the actomyosin meshwork of animal cells,
are ubiquitous in living organisms. The active gel theory,
which provides a thermodynamic framework for these materials,
has been mostly used in conjunction with the assumption that
the microstructure of the biopolymer network is based on rigid rods.
However, experimentally, {crossed-linked} actin networks exhibit entropic elasticity. Here we combine an entropic elasticity kinetic theory, in the spirit
of the Green and Tobolsky model of transiently crosslinked networks,
with an active flux modelling biological activity. We determine this
active flux by applying Onsager reciprocal relations to the corresponding microscopic dynamics. We derive the macroscopic active stress that arises from the resulting dynamics and obtain a closed-form model of the macroscopic mechanical behaviour. We show how this model can be rewritten using the framework of multiplicative deformation gradient decomposition, which is convenient for the resolution of such problems.
\end{abstract}

\section{Introduction}
\label{sec-intro}

Active matter comprises a wide range of structures of mechanical relevance
and features numerous means to perform mechanical work with and within them \cite{Taber.1995.1}.
Among those, within the animal kingdom, biopolymer networks endowed
with the ability to self-contract are arguably the most pervasive, both inside
the cells and outside of them.
Muscle contraction constitutes a classical example of this. It can
be idealised as the relative sliding of parallelly-organised filaments of actin and myosin,
 actuated by the conformation change of the myosin, which itself is powered by the energy
released by ATP hydrolysis \cite{Huxley.1957.1,Caruel-Truskinovsky.2018.1}.
Actin and myosin are also essential players of the cytoskeleton. They are involved in an extensive range of active
mechanical behaviours of cells \cite{Bray-White.1988.1,Mitchison-Cramer.1996.1,Recho+.2013.1}, where they form a
thin crosslinked network apposed to the plasma membrane 
\cite{Salbreux+Paluch.2012.1}, called the \emph{actomyosin} cortex. 
%Outside of cells, extracellular matrices (ECM) are based on networks of biological polymers
%that form soft elastic gels \cite{Levental+Janmey.2007.1} and within which cells exert forces in
%a variety of ways \cite{Delvoye+Lapiere.1991.1,Bertinetti+Fratzl.2015.1,Han+Guo.2018.1}.
The very rich repertoire of contractile networks has also been explored in vitro using purified
gels of biological proteins \cite{Nedelec+Leibler.1997.1,Koenderink+Weitz.2009.1,Stuhrmann+Koenderink.2012.1}.
While %ECM networks and 
in vitro gels have properties that are distinct from those of cytoskeletal
actomyosin, the geometry of the microstructure and the origins of active stress bear interesting similarities.

Continuum modelling approaches have been able to reproduce a number of the behaviours
of the actomyosin cytoskeleton by the introduction of an active stress as a driving force within a liquid-like
material modelling the network \cite{Dembo-Harlow.1986.1}.
This active stress can be interpreted as a dynamic prestress \cite{Erlich+Wyatt.2022.1}, allowing
to draw analogies with the residual stress which is observed in solid-like
tissue \cite{Fung.1991.1,Goriely-Vandiver.2010.1}.
The theory of active gels \cite{Kruse+.2005.1,Juelicher+.2007.1}, drawing from the hydrodynamics
of suspensions of orientable objects endowed with active stresses \cite{Marchetti+Simha.2013.1},
has provided a sound thermodynamic framework for the generation of this active stress by molecular
motors.
A link has been done with materials whose microstructure relies on 
rigid filaments \cite{Liverpool-Marchetti.2006.1,Ahmadi+Liverpool.2006.1,Hawkins-Liverpool.2014.1}.  
Actin filaments are semiflexible, however, {to the difference of lamellipodium-like networks \cite{Pujol+Heuvingh.2012.1},} the elasticity of crossed-linked networks has
been shown to be of entropic nature \cite{Gardel+Weitz.2004.2}. It is thus interesting to
consider in what measure the entropic elasticity of the microstructure modifies the
constitutive relations of active biopolymer networks.

Here, we consider a microstructure of freely-jointed chains, exhibiting entropic elasticity,
which form a percolating network through high-affinity but reversible binding. 
In addition to affine deformations, we allow for motion due to the action of active crosslinks,
which model molecular motors. The thermodynamics of this system is then written within the
constraints of this particular microstructure, which allows us to propose microscopic interpretations
of the Onsager relations. 

Section \ref{sec-kinetic} presents the kinetic model, its thermodynamics and is concluded with
a closed mechanical model. Section \ref{sec-multiplicative} introduces a multiplicative decomposition
that allows to recover this model and is convenient for its resolution and for numerical
approaches. Section \ref{sec-examples} concludes the paper with two simple examples demonstrating the method.

\section{A kinetic theory of contractile biopolymer networks}
\label{sec-kinetic}
\linespread{1.1}

\subsection{Kinetics of active temporary networks}
\label{sec-smolu}

In this section, we follow the approach of transiently crosslinked networks \cite{Green-Tobolsky.1946.1,Yamamoto.1956.1}
and combine it explicitly with an 
elastic dumbbell model \cite{Bird+.1987.2,Larson.1988.1,Larson.1999.1} to describe the dynamics
of unbound chains. This will make possible the interpretation
of the dissipation in the next section. The other novelty here is the presence of an additional flux,
which represents the active dynamics of the network. It is not specified in this section how this
flux depends on other fields, this will be done in view of the thermodynamics of the system in 
\refsec{sec-motors}, allowing us to reconcile phenomenological kinetic approaches \cite{Etienne+Asnacios.2015.1}
with generic thermodynamic considerations \cite{Kruse+.2005.1}.

\newlength{\figw}
\setlength{\figw}{\textwidth}
\begin{figure}
\begin{center}
A
\includegraphics[width=0.55\figw]{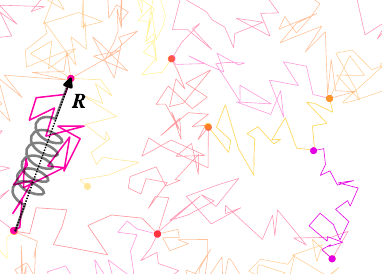}
\vspace{5mm}\\
B
\includegraphics[width=0.55\figw]{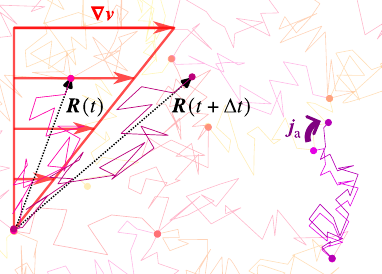}
\vspace{5mm}\\
C
\includegraphics[width=0.55\figw]{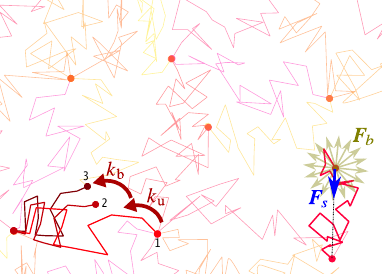}
\end{center}
\caption{
Network model. \textbf{A}, Chains made of a finite  number of freely-jointed segments. Each
is described  by an end-to-end vector $\vc{R}$ and can be modelled as a spring. 
They can connect to other chains at their ends.
\textbf{B}, If a global deformation $\grad\vc{v}$ exists, bound chains
are affinely deformed, \eq{eq-langevin-b}.
The additional active flux $\Ja$ is described in \refsec{sec-motors}.
\textbf{C}, Unbound end of a chain is submitted to spring force $\vc{F}_s$ 
and to a Brownian force $\vc{F}_b$.
Unbound chains relax extremely fast to a configuration where those forces balance.
Bound ends (configuration 1) unbind at rate $\ku$ and the chain thus relaxes (configuration 2).
They will then quickly rebind at rate $\kb$ (configuration 3) but do so in their relaxed state.
}
\label{fig-network}
\end{figure}

We consider the biopolymer network as an assembly of polymer molecules of idealised elastic
freely-jointed chains in a viscous liquid (e.g. the cytosol). Each chains can be represented by an elastic
dumbbell \cite{Larson.1999.1}, that is, a spring joining two beads which can bind to beads of neighbouring
molecules. For each chain, the end-to-end vector $\vc{R}(t) \in \Reals^3$, called \emph{strand}, connects each of the 
dumbbell beads, see \fig{fig-network}.

We assume a high affinity between the chains and that
consequently, a large connected component of
chains forms a macroscopic network. Within this percolated component, each
"bound" strand $\Rb$
is assumed to deform affinely with the network. Thus, if the macroscopic
velocity gradient $\tsL\,\trsp$ varies over distances much greater than the
typical length of $\Rb$, the dynamic of $\Rb$ is given by :
\begin{align}
\dRb =\tsL\cdot \Rb + \Ja.
\label{eq-langevin-b}
\end{align}
Here $\Ja$ represents some active process that
remodels the network, yet unspecified, the case of 
passive transiently crosslinked networks being $\Ja=\vc{0}$.

However, transiently, one of the beads in a dumbbell can 
disconnect from this network, undergo different dynamics, and then reconnect to it.
\JEadds{
The rate constants governing connection and disconnection will be set such that these
transients are of short duration, one consequence being that 
we do not need to consider the quadratically rare cases where both beads disconnect.
When one of the beads is disconnected, it is subjected to the internal spring force $\vc{F}_s$ exerted by the
chain, which remains bound to the network by the other bead, and to the Brownian force $\vc{F}_b$ 
due to the thermal fluctuations in the liquid.
This can be expressed in a stochastic differential equation, called a 
Langevin equation, which governs the dynamics of the "unbound" strand $\Ru$:
}
\begin{align}
\zeta(\dRu - \tsL\cdot\Ru) = \vc{F}_s + \vc{F}_b,
\label{eq-langevin-u}
\end{align}
\JEadds{where $\zeta$ is the drag coefficient of the bead in the fluid.}
The liquid is considered to have the
same velocity locally as the network, thus the viscous drag force on the free end of the chain
is proportional to its velocity relative to $\tsL\cdot\Ru$,
where we have again considered that the variations of $\tsL$ are on a larger spatial scale than
the typical $|\Ru|$.

Up to a factor two on the drag coefficient $\zeta$ (due to the bound state of one of the beads), 
this is the same Langevin equation as the
one used to derive dumbbell models \cite{Bird+.1987.2,Larson.1988.1,Larson.1999.1}, 
and the expressions of the forces are the same:
\begin{align}
\vc{F}_b &= \sqrt{2\zeta k_BT} \vc{\eta}(t),\\
\vc{F}_s &= - \kappa \vc{R}_u.
\end{align}
The form of $\vc{F}_b$, with $k_B$ the Boltzmann constant, $T$ the temperature and $\vc{\eta}$ an isotropic white noise  is chosen to guarantee the equipartition of energy in the permanent regime
\cite{Larson.1988.1} for the (passive) unbound phase.
The strand stiffness $\kappa = 2k_BT\beta^2$ is normalized with  the thermal energy  where $\beta^{-1}$ is a length proportional to the square root of the number of segments in the chain.

\JEadds{We now introduce} the probability densities for a strand to be bound (resp.\  unbound) and of a certain length $\psi_{\mathrm{b,u}}(\vc{r},t)\mathrm{d}\pt{r} = \Prob\{\text{b,u}\}\cdot \Prob(r_i<\vc{R}(t)\cdot\vc{e}_i<r_i+\mathrm{d}r_i)$.
The Fokker-Planck dynamics associated to the Langevin process \eqref{eq-langevin-b}-\eqref{eq-langevin-u} \JEadds{are described by
a conservation equation for each of these
probability densities, taking into account the `probability current' defined by the Langevin process and the exchanges between the two states, bound and 
unbound. They} take the form:
\begin{subequations}
\label{eq-smolu-both}
\begin{align}
\dedt{\psib} + \gradr\cdot\left(\psib(\tsL\cdot \vc{r} + \Ja)\right) &= -\ku\mathcal{K}_u(\psiu,\psib)
\label{eq-smolu-psib}
\\
\dedt{\psiu} 
 + \gradr\cdot\left(
     \psiu\big(\tsL\cdot \vc{r}- (\kappa/\zeta) \vc{r} \big)
     -(k_BT/\zeta)\gradr \psiu
 \right) &= \hphantom{-}\ku\mathcal{K}_u(\psiu,\psib)
\label{eq-smolu-psiu}
\end{align}
\end{subequations}
where \textcolor{red}{the left hand sides are obtained in the absence of exchange terms between the bound and unbound states \cite{risken1996fokker} and a reaction term $\ku\mathcal{K}_u$ appears on the right hand side in order to model binding and unbinding. The  characteristic time for this process is the unbinding rate $\ku$}. In the sequel 
we will take the simplest possible reaction term,
$\mathcal{K}_u = \JEadds{ \psib - (\kb/\ku)\psiu}$, 
with $\kb$ a constant binding rate. Note that the total probability distribution of the two populations,  $\psi = \psiu + \psib$,
is such that $\intd{\Reals^3}{}{\, \psi }{\pt{r}} = 1$. 

We now consider the limit of low drag resistance to the motion of unbound chains: this
corresponds to a fast relaxation to equilibrium, \JEadds{happening before a chain rebinds to}
the network, and thus $\zeta \kb/\kappa \ll 1$.
\JEadds{As already stated, the unbinding process is assumed to be slower than the binding one, $\ku \lesssim \kb$, which is necessary to obtain a large connected  component forming the network. We additionally restrict ourselves to rates of flow which cannot exceed the binding rate, $|\tsL|\lesssim k_b$, to prevent mechanical rupture. Under these conditions,}
we can then expand $\psiu$ as $\psiu = \au\psi_0 + O(\zeta \kb/\kappa)$
where $\psi_0(\vc{r})$ is time-independent and solves \eq{eq-smolu-psiu} to the first order, see \JEadds{\refapp{app-unbound}. 
The fraction of bound chains $\au$ is found to tend to a permanent regime value characterised by $\intd{\Reals^3~}{}{\mathcal{K}_u(\psiu,\psib)}{\pt{r}}={0}$. 
For our choice of $\mathcal{K}_u$, this results in $\au = \ku/(\kb+\ku)$.
%The asymptotics chosen here are such that $\max\{ |\tsL|, \ku \} \lesssim \kb \ll \kappa/\zeta$, thus the shear rate and unbinding rate need to be lower than the binding and chain relaxation rates. 
We will often consider the case $\ku \ll \kb$, since this simplifies some of the expressions.%, but the shear rate does not have to be small relative to its upper bound $\kb$ and can in particular be large compared to $\ku$, which will be shown below to be the relaxation rate of the system. 
}
\hide{
The distribution $\psi_0$
corresponds to the Gaussian distribution that freely-jointed
chains adopt at rest \JE[for $\vc{F}_s+\vc{F}_b=0$]
{avec le $F_b$ stochastique ça ne marche plus, et de même pour l'eq en annexe A, à moins de définir un $F_b$ moyenné...?}
, leading to $-\psi_0 \kappa \vc{r} -k_BT\gradr \psi_0=0$ and thus
$\psi_0 = (\kappa/(2\pi k_BT))^{3/2} \exp(-\kappa \pt{r}^2/(2k_BT))$
\JEadds{where $\pt{r}^2=\vc{r}\cdot\vc{r}$}.}

The time dependent part of $\psi$ is thus reduced to $\psib$, and from
\eq{eq-smolu-psib} we have:
\begin{align}
\dedt{\psib} + \gradr\cdot\left(\psib(\tsL\cdot\vc{r} + \Ja)\right) 
	&= -\ku\mathcal{K}_0(\psib) + O\left(\frac{\zeta \kb\ku}{\kappa}\right)
\label{eq-smolu}
\end{align}

For $\Ja=\vc{0}$
and since the first order approximation is
$\ku\mathcal{K}_0(\psib)=\ku\psib - \kb \au \psi_0$,
\eq{eq-smolu} is the same Smoluchowski equation as found by \citetext{Yamamoto.1956.1} for $\psib$,
this is known to lead to the upper-convected Maxwell 
constitutive equation for stress--strain relation
\cite[p. 168]{Larson.1988.1}, 
with a relaxation time equal to $\ku^{-1}$. Note that the upper-convected
Maxwell stress--strain relation, with relaxation time equal to
$\zeta/\kappa$,  is also 
the result for unbound chains only, i.e.\ \eqs{eq-smolu-psiu},
with $\mathcal{K}_u=0$ but finite $\zeta$.

\subsection{The stress tensor}
\label{sec-stress}

\JEadds{Elastic forces in stretched polymer chains result in a stress in the network, which can be described using the distribution $\psi$ and the spring force $\vc{F}_s$.
}%
The classical derivation \JEadds{of the stress tensor} uses the procedure of integrating the forces exerted by this distribution of springs on an
elementary volume, but the same result can be obtained using the virtual work principle \cite[p. 33]{Larson.1988.1}.
As will be seen below, this is useful in the context of active systems.

For our isothermal system, the total dissipation can be written as:
$$
\mathcal{D} = \frac{\delta W}{\delta t} - \ddt{\mathcal{F}} \geq 0
$$
where the terms are respectively the rate of work and the Gibbs free energy variations.
\JEadds{The system is actually open, since an energy input from the outside maintains out of equilibrium a chemical reaction feeding the power strokes of molecular motors  \cite{DESHPANDE2021104381}, which are modelled with the term $\Ja$. When this reaction is maintained at a fixed distance from equilibrium, we include the energy exchange term with the environment in $\mathcal{F}$, as classically done in the active gel theory \cite{Kruse+.2005.1}. This will be specified in \refsec{sec-motors}.}

\JEadds{We now proceed to determine the rate of work.}
Let $\Omega(t)$ be the domain occupied by the material at time $t$.
The trajectory of material points of the network is
denoted $\vc{x} = \vc{\chi}(\vc{X},t) \in \Omega(t)$ where $\vc{X}\in \Omega(0)$ is the initial  position at time $t=0$.
We define the deformation \JEadds{gradient as $\ts{F}(\vc{X},t) = \grad_{\!\pt{X}} \vc{\chi}(\vc{X},t)\trsp$,
with the convention $\grad_{\!\pt{X}} \vc{U} =\sum_{i,j} \partial U_j/\partial X_i \vc{e}_i\vc{e}_j$. 
We follow the conventions of \citetext{Larson.1999.1} among others, notably
the outer (dyadic) product is implied between vectors, $\vc{U}\vc{V} = \sum_{i,j} U_i V_j \vc{e}_i\vc{e}_j$ 
while $\,\cdot\,$ denotes the scalar product of vectors, the tensor--vector product and the contracted product of
second order tensors $\ts{T}\cdot\ts{S} = \sum_{i,j,k} T_{i,k} S_{k,j} \vc{e}_i\vc{e}_j$. Finally, 
$\,:\,$ is the doubly contracted product, $\ts{T}:\ts{S} = \sum_{i,j} T_{i,j} S_{j,i}$.}

The velocity is 
$$\vc{v} = \dedt{\vc{\chi}}(\vc{\chi}^{-1}(\vc{x},t),t),$$
and we identify
$\tsL=\gradx\vc{v}\trsp$. To remain in the conditions above, it is assumed that $\ts{F}(\vc{X},t)$
varies in space over distances much longer than the typical length of a strand $\vc{R}$.
We assume that inertia and body forces acting on the system are negligible,
leading to the momentum balance equations:
\begin{subequations}
\label{eq-momentum}
\begin{align}
%-\grad p  + \div\tsigma 
\gradx\cdot\tsTau &= \vc{0} 	& \text{in }\Omega(t)
\\
%-p\vc{n} + \tsigma
\tsTau \cdot \vc{n} &= \vc{f}_\text{ext} 	& \text{on }\partial\Omega(t)
\label{eq-bc-force}
\end{align}
\end{subequations}
%\PR{Manque un div à $\tau$ mais je ne sais pas comment tu veux l'écrire}
where $\vc{f}_\text{ext}$ are the external forces applied on the system's boundaries
%\JEadds{and we have decomposed the stress term into $p$, the %\JE[fluid]{à moduler si...} pressure and $\tsigma$, the so-called \emph{extra stress tensor} \cite{Larson.1999.1}.}
and \JEadds{$\tsTau$ the Cauchy stress tensor.}

Let us denote $\nu(\vc{x})$ the number of chains per unit volume within $\Omega(t)$.
Mass conservation gives
\begin{subequations}
\label{pb-mass_csv}
\begin{align}
\dedt{\nu} + \gradx\cdot(\nu \vc{v}) 	&= 0
	& \text{in }\Omega(t)
\label{eq-mass_csv}
\\
\nu ( \vc{v} - \vc{v}_b ) \cdot \vc{n} &= 0 
	& \text{on }\partial\Omega(t)
\\
\nu(t=0) &= \nu_0			     	& \text{in }\Omega_0 = \Omega(0)	
\end{align}
\end{subequations}
where $\vc{v}_b(t)$ is the velocity of $\partial\Omega(t)$, which from the above has its normal component equal
to $\vc{v}$.

The rate of work is equal to the work performed on the
system's boundary, and from \eq{eq-momentum} and using the symmetry of $\tsTau$, %$\tsigma$,
\begin{align}
\frac{\delta W}{\delta t} = \intd{\partial\Omega(t)}{}{ \vc{f}_\text{ext} \cdot \JEadds{\vc{v}_b} }{\pt{\pt{s}}}
	= \intd{\Omega(t)} {}{\tsTau:\tsD \,}{\pt{x}}
	\label{eq-energyvar-work}
\end{align}
with $\tsD = \frac{1}{2}(\tsL+\tsL\trsp)$.

\hide{
Following \cite[p. 33]{Larson.1988.1}, the elastic  contribution of $\nu$ chains per
unit volume to the Gibbs free energy is
\begin{align*}
\nu\varphi_{\mathrm{e}} :=& \frac{1}{2}\nu\kappa\intd{\Reals^3}{}{\psi(\vc{r})\pt{r}^2}{\pt{r}}.
\end{align*}}
We now write the Gibbs free energy of the whole system: 
$$
{\mathcal{F}} = \intd{\Omega(t)}{}{\nu\varphi}{\pt{x}}.
$$
The specific free energy density $\varphi$ can be decomposed as $\varphi = \varphi_{\mathrm{e}} + \varphi_{\mathrm{a}} $, where $\varphi_{\mathrm{a}}$ represents anelastic contributions, which we will relate to active processes in \refsec{sec-motors},
and $\varphi_{\mathrm{e}}$ is the specific elastic free energy density.
Following \cite[p. 33]{Larson.1988.1}, %the elastic  contribution to the Gibbs free energy %density of $\nu$ chains per unit volume with a distribution $\psi$ is
it is:
\begin{align*}
\nu\varphi_{\mathrm{e}} :=& \frac{1}{2}\nu\kappa\intd{\Reals^3}{}{\psi(\vc{r})\pt{r}^2}{\pt{r}}.
\end{align*}
\JEadds{where $\pt{r}^2=\vc{r}\cdot\vc{r}$.}

The free energy variations are:
\begin{align}
\ddt{\mathcal{F}} &= \intd{\Omega(t)}{}{\left(  \dedt{\nu} \varphi + \nu \dedt{\varphi} \right) }{\pt{x}}
	= \intd{\Omega(t)}{}{
		\nu \ddt{\varphi} 
		}{\pt{x}}
	\label{eq-energyvar-Gibbs}
\end{align}
where we have used \eq{eq-mass_csv} and one integration by parts.

Since in the limit $\lfrac{\zeta\kb}{\kappa}\rightarrow 0$, $\psiu$ is time-independent, the elastic contributions to free energy density variations
can be deduced from the Smoluchowski \eq{eq-smolu}:
\begin{align*}
\nu\ddt{\varphi_{\mathrm{e}}} 
	&= \frac{1}{2}\nu \kappa \intd{\Reals^3}{}{\dedt{\psib}\pt{r}^2}{\pt{r}}
	\\
	&\stackrel{\text{IbP}}{=} \frac{1}{2}\nu\kappa\intd{\Reals^3}{}{\psib (\vc{r}\cdot\gradx\vc{v}+\Ja) \cdot\gradr\pt{r}^2}{\pt{r}}
		- \frac{1}{2}\nu \kappa \ku \intd{\Reals^3}{}{\mathcal{K}_0(\psib) \pt{r}^2}{\pt{r}} 
	\\
	&=  \nu\kappa\left(
		\langle\vc{r}\vc{r}\rangleb:\tsD
		+ \langle\Ja\cdot\vc{r}\rangleb
		- \frac{1}{2}\ku \intd{\Reals^3}{}{\mathcal{K}_0(\psib) \pt{r}^2}{\pt{r}}
	    \right)
\end{align*}
where $\langle\,\cdot\,\rangleb = \intd{\Reals^3~}{}{\psib \,\cdot\,}{\pt{r}}$.
\JEadds{It is thus the second moment of the distribution of bound strands, 
$\tA = \langle\vc{r}\vc{r}\rangleb$, which is conjugated with the rate of strain $\tsD$.
This tensor $\tA$ is called the \emph{microstructure (or texture) tensor} \cite{Aubouy+Graner.2003.1}
of the network. 
}

We are now in a position to identify the different terms in the dissipation,
\begin{align*}
\mathcal{D} 
 &= \mathcal{D}_{\mathrm{w}} + \mathcal{D}_{\mathrm{r}} + \mathcal{D}_{\mathrm{a}}, %	- \ddt{\mathcal{F}_{\!\mathrm{a}}}
\end{align*}
where the first term is a dissipation induced by the deformation rate:
\begin{align*}
\mathcal{D}_{\mathrm{w}} &= \intd{\Omega(t)}{}{ \left(\tsTau-\nu\kappa\tA\right):\tsD }{\pt{x}},
\intertext{the second term is a dissipation associated with the relaxation of chains via the unbinding--rebinding dynamics:}
\mathcal{D}_{\mathrm{r}} &= \JEadds{\frac{1}{2}}	 \intd{\Omega(t)\,}{}{\nu \kappa \ku\intd{\Reals^3}{}{\mathcal{K}_0(\psib) \pt{r}^2}{\pt{r}}  }{\pt{x}}
\intertext{and the third is the power balance of active processes:}
\mathcal{D}_{\mathrm{a}} %-\ddt{\mathcal{F}_{\!\mathrm{a}}}
	&= -\intd{\Omega(t)}{}{   
		 \nu\left( \ddt{\varphi_{\mathrm{a}}}
		+ \kappa\langle\Ja\cdot\vc{r}\rangleb
		\right)}{\pt{x}}.
\end{align*}

One can further decompose the stress term into deviatoric and isotropic components,
$$
\mathcal{D}_{\mathrm{w}} 
 = \intd{\Omega(t)}{}{\left(\dev(\tsTau-\nu\kappa\tA):\tsD 
	  + \frac{1}{3}\tr(\tsTau -\nu\kappa\tA) \ts{I}:\tsD
	\right) 
	}{\pt{x}}.
$$
\JEadds{In what follows we choose to treat the case of an incompressible
material only, which yields $\ts{I}:\tsD=\div\vc{v}=0$ and
cancels the second term of the
integrand. This is enforced by a
pressure $p$ as a Lagrange multiplier.
Following e.g. \cite{Larson.1999.1}, we decompose the stress $\tsTau = -p\ts{I} + \tsigma$ where $\tsigma$ is the so-called \emph{extra stress tensor}.}
In order to guarantee $\mathcal{D}_{\mathrm{w}}\geq 0$, 
the extra stress tensor can then be chosen as 
$\tsigma=\nu\kappa(\tA-\tA_{0}) + 2\mu\tsD$, % +\mu_b(\grad\cdot\vc{v})\ts{I}$, 
%where $\mu_b$ and $\mu$ are Lamé parameters corresponding to the dissipative
where $\mu$ introduces an additional dissipation not related with the microstructure,
typically ascribed to a shear viscosity of the liquid bath. 
\JEadds{Incompressibility implies that isotropic terms in the stress are not 
contributing to the dissipation.}
We have thus introduced an arbitrary isotropic tensor $\tA_0=a_0\beta^{-2}\ts{I}$, 
\JEadds{and we will see in the next section that the relaxed state of the chains prescribes
a nonzero $a_0$.}
%where the
%value of $a_0$, corresponding to relaxed chains, will be derived in the next
%section. 
%Its contribution to the stress can be balanced by the pressure term, indeed, the pressure $p$ can be chosen as a
%Lagrange multiplier which will enforce $\ts{I}:\tsD=\div\vc{v}=0$ or any prescribed volume change.
%\JE[For both of these constitutive choices,]{...not updated... and then below no $\mu_b$...}
We obtain 
$$
\mathcal{D}_{\mathrm{w}} = \intd{\Omega(t)}{}{ 2\mu \tsD:\tsD }{\pt{x}} \geq 0,
%\mathcal{D}_w = \intd{\Omega(t)}{}{\left(2\mu \tsD:\tsD + %\mu_b(\div\vc{v})^2\right)}{\pt{x}},
$$
thus only the liquid bath may contribute to a dissipation induced by the rate of deformation.

\subsection{Active terms: molecular motors as crosslinks}
\label{sec-motors}

We now turn to the modelling of the molecular motors. 
We have assumed that their effect was felt through a flux $\Ja$ 
in the dynamics of bound chains, \eq{eq-langevin-b}, since indeed
myosin minifilaments are a crosslinking type of molecular motors which can actively 
displace their binding position along either of the actin filaments they
are bound to during events called power-strokes \cite{Lipowsky-Liepelt.2008.1}.
We now investigate what form of $\Ja$ is admissible from a close-to-equilibrium thermodynamics point of view.

The power-stroke process is driven by a chemical reaction in which ATP hydrolysis into ADP
releases mechanical energy in myosin motor heads.
The internal energy that fuels this reaction can be written as
the product of the average advancement of the reaction $\langle\xic\rangleb$ 
and an affinity 
$\Deltamu$, $\varphi_{\mathrm{a}}=-\Deltamu \langle\xic\rangleb$.
Following \cite{Kruse+.2005.1}, we consider that $\Deltamu$ is maintained
a constant throughout the process.

We now resort to an Onsager approach to determine the active flux $\Ja$
%which gives rise to the microstructure active deformation rate tensor $\ku\tAa$, 
specifying the power balance of the molecular motors,
\begin{align*}
\mathcal{D}_{\mathrm{a}} %\ddt{\mathcal{F}_{\!\mathrm{a}}}
	&= \intd{\Omega(t)}{}{
	    \nu\left( 
	      \Deltamu \langle\dxic\rangleb
	    - \kappa\langle\Ja\cdot\vc{r}\rangleb  
		\right)}{\pt{x}}.
\end{align*}

Since %$\intd{\Reals^3~}{}{\psib}{\pt{r}}=1-\au=1+O(\ku/\kb)$ and since 
$\vc{F}_s = -\kappa \vc{r}$,
\begin{align*}
\Deltamu \JEadds{\langle\dxic\rangleb} - \kappa\langle\Ja\cdot\vc{r}\rangleb 
  &=  \intd{\Reals^3~}{}{ \left(  \Deltamu\dxic + \vc{F}_s\cdot\Ja  \right)\psib}{\pt{r}}
\end{align*}
making apparent the pairs of conjugate fluxes and forces $(\dxic,\,\Deltamu)$
and $(\Ja,\,\vc{F}_s)$. 
\JEadds{In the same way as \cite{Julicher+Prost.1997.1}, 
we thus expand linearly the fluxes as}:
\begin{subequations}
\label{eq-onsager}
\begin{align}
\Ja &= \lambda_{11} \vc{F}_s + \vc{\lambda}_{12}\Deltamu ,
\\
\dxic &= \vc{\lambda}_{21} \cdot \vc{F}_s + \lambda_{22} \Deltamu.
\end{align}
\end{subequations}

The coefficient $\lambda_{11}\geq 0$ describes a slippage friction of the bound crosslinkers with respect to the polymer strand,     
which may constitute an additional source of relaxation of the network along with the unbinding--rebinding
process. We can characterise it with a rate $\frac{1}{2}\ks$ and a typical energy of the strands $\kappa \beta^{-2}$,
yielding $\lambda_{11}=\ks /(2\kappa)$.
The other dissipative coefficient $\lambda_{22}\geq 0$
corresponds to the dissipation in the chemical reaction itself. As the dissipation $\mathcal{D}_a$ is positive, 
\begin{equation}\label{e:thermo_ineq}
 \lambda_{22}\lambda_{11}\geq \vc{\lambda}_{21} \cdot \vc{\lambda}_{12}.
\end{equation}
Following Onsager symmetry relations \cite{deGroot-Mazur.1984.1},
we take equal reactive coefficients
$\vc{\lambda}_{12}=\vc{\lambda}_{21}$.
Since they are vectorial, a vector quantity has to be 
constructed at the microscale. One possibility is to use the strand vector
$\vc{r}$ itself, which corresponds to the
assumption that the orientation of $\vc{r}$ has a microscopic relevance.
This is true for actin filaments, which are oriented, and myosin molecular
motors which are able to sense this orientation. We call this the
processive flux, use again $\kappa\beta^{-2}$ as the typical strand elastic 
energy to normalise the coefficients,
\begin{align*}
\vc{\lambda}_{12} &= \frac{\va\beta^{3}}{\kappa} \theta(\pt{r}) \vc{r},
&
\Ja &= 
  \left(
  	\frac{\va \beta^3\Deltamu}{\kappa} \theta(\pt{r}) -\frac{1}{2}\ks  
  \right)
  \vc{r},
\\
&
&
\dxic &= -\va{\beta^3}\theta(\pt{r})\pt{r}^2 + \lambda_{22} \Deltamu,
\end{align*}
where $\theta$ is some nondimensional function of $\vc{r}$ \JEadds{and  we have introduced the velocity of processivity of motors $\va$, which, in the absence of slippage, can be related to the rate of detachment $\ku$ by the distance $\ell_{\mathrm{a}}$ that motors
travel along a strand before detaching, $\ell_{\mathrm{a}} = \va/\ku$. If there is a nonzero slippage rate $\ks$, we can interpret a distance $\ell_{\mathrm{a}} = \va/(\ku+\ks)$ in a similar way.}
%The microstructure active deformation rate tensor that arises is then characterised as:
\JEadds{The specific energetic contribution of the flux $\Ja$ can then be characterised as proportional to:}
\begin{align}
\JEadds{\langle \vc{r} \cdot \Ja\rangleb} &=   \frac{\ku+\ks}{2} \tAa:\ts{I} - \frac{\ks}{2}\ts{A}:\ts{I}
&&\text{with}&
\tAa &= \frac{2\ell_{\mathrm{a}}\beta^3\Deltamu}{\kappa} \langle\theta(\pt{r})\vc{r}\vc{r}\rangleb.
\label{eq-contractilitytensor}
\end{align}
The slippage introduced by $\ks$ thus results in a dissipation,
whereas the first term is the trace of what we define as a \emph{contractility tensor} $\tAa$. We already define an associated \emph{active stress tensor} $\tsiga = \nu\kappa\tAa$, which will be useful in \refsec{sec-veac}.
%Without loss of generality we will take $\ks=0$ in what follows.

One interesting case is %to require that the rate of advancement of the reaction is
%independent of the mechanical stress in the filament. This leads to 
%The constraint of $\mathcal{D}_a>0$ is met for $\lambda_22 > \frac{\pt{\lambda}_21^2}{\lambda_11}$,
to take $\theta(\vc{r}) = 1/({\beta\pt{r})^2}$.
%and for simplicity we can define a typical length $\ell_{\mathrm{a}}$ which motors %travel 
%before a disconnection event, $\ka = \ku \ell_{\mathrm{a}}\beta$.
%This introduces a stalling force behaviour for the motors. 
Indeed, we then 
have
$$
\Ja^{\text{0}} = (\ku+\ks)\ell_{\mathrm{a}}\frac{\beta\Deltamu}{\kappa|\vc{r}|} \frac{\vc{r}}{|\vc{r}|}
$$
where the motors follow the filament direction ${\vc{r}}/{|\vc{r}|}$ but
proceed with a velocity that decreases hyperbolically with increasing strand tension 
$\kappa|\vc{r}|$. 
%We have introduced the typical length $\ell_{\mathrm{a}}$ which motors travel before a disconnection event, setting \JE[$\ka = \ku \ell_{\mathrm{a}}\beta$.]{adjust}
The microstructure contractility tensor and active stress that arise can then be explicited as:
\begin{align*}
\tAa &= \frac{2\ell_{\mathrm{a}}\beta\Deltamu}{\kappa} \langle\ts{Q}\rangleb,
&
\tsiga &={2\nu\ell_{\mathrm{a}}\beta\Deltamu} \langle\ts{Q}\rangleb.
\end{align*}
with $\langle\ts{Q}\rangleb=\langle\vc{r}\vc{r}/\pt{r}^2\rangleb$ the orientation tensor of the network, and is seen,  for this particular choice of $\theta$, to be independent of how much the network is stretched.
Note that the `stalling' behaviour of the motors in $\Ja^{\text{0}}$ does not explicitly appear
at the macroscopic scale, and conversely we have shown previously that macroscopic stalling can be
a collective effect which is only modulated by molecular-scale stalling \cite{Etienne+Asnacios.2015.1}.
\JEadds{At the microscopic level, the thermodynamic inequality \eqref{e:thermo_ineq} on $\lambda_{22}$ implies that this choice of $\theta(\pt{r})$ is only valid if the network can be shown not to collapse, $|\pt{r}|\geq r_0>0$.}

\JEadds{%Under the condition (\ref{eq-lambdatoutou}), 
The total contribution of active terms to dissipation is:}
\begin{align*}
\mathcal{D}_{\mathrm{a}}
	&= \intd{\Omega(t)}{}{ \nu(\ks\tA:\ts{I} \JEadds{-2{\va\beta^3\Deltamu}} + \lambda_{22}\Deltamu^2) }{\pt{x}} %\geq 0.
\end{align*}
\JEadds{and is positive under the condition that:
\begin{align}
\lambda_{22} \geq 
\frac{2(\ku+\ks)^2 \ell_{\mathrm{a}}^2}{\ks k_BT} (2\pi\beta^{-1}\|\psib\|_{\infty}+\beta^2) \geq
 \frac{2(\ku+\ks)^2 \ell_{\mathrm{a}}^2}{\ks k_BT} \left\langle\frac{1}{\pt{r}^2}\right\rangleb.
\label{eq-lambdatoutou}
\end{align}
In what follows, we occasionally take the limit $\ks\ll\ku$ whereas $\ku$ is close to the characteristic time of the processes of interest, implying that $(\ell_{\mathrm{a}}\beta)^2$ is small, limiting in turn the magnitude of $\tsiga$. Note however that this is only for comparison with passive systems, since it simplifies the expression of factors, but it is in no way necessary.}

Other choices are possible for the reactive coefficient $\vc{\lambda}_{12}$, see \refapp{app-motors}. In particular,
a model of ``diffusive'' flux along the filaments leads to the same
form of the contractility tensor and active stress but a different prefactor.
The orientation tensor $\langle\ts{Q}\rangleb=\langle\vc{r}\vc{r}/\pt{r}^2\rangleb$ can be
seen as a generalisation of the nematic tensor, in that it allows e.g. isotropy in 
a plane tangential to a surface.

%\JEadds{%Regardless of the choice of $\vc{\lambda}_{12}$, 
%Thus, the contribution of active terms to dissipation is:}
%\begin{align*}
%\mathcal{D}_{\mathrm{a}}
%	&= \intd{\Omega(t)}{}{ \nu(\ks\tA:\ts{I} \JEadds{-2{\va\beta^3\Deltamu}} + \lambda_{22}\Deltamu^2) }{\pt{x}} %>0.
%\end{align*}
%\JEadds{Since $\tA = \langle \vc{r}\vc{r}\rangleb$, its trace is always %positive which guarantees the dissipative nature
%of this term.}

Note that the velocity $\va$ and corresponding length $\ell_{\mathrm{a}}$ can depend on the position 
in physical space, e.g.\ they may depend on the local concentration of some catalytic species.

\subsection{Constitutive equation of active networks}
\label{sec-veac}

Having defined the stress tensor in \refsec{sec-stress}, we can proceed with the usual procedure of bead--spring models
in order to determine the constitutive equation that relates stress and strain, and which consists 
in multiplying the Smoluchowski \eq{eq-smolu} by the tensor $\vc{r}\vc{r}$ and integrating over the
phase space. This yields:
\begin{align}
\ddt{\tA} -\tsL\cdot\tA-\tA\cdot\tsL\,\trsp = -\ku \intd{\Reals^3}{}{\mathcal{K}_0(\psib) \vc{r}\vc{r}}{\pt{r}} 
+ \langle \vc{r}\Ja+\Ja\vc{r} \rangleb %which is + \ka \tAa 
\label{eq-relaxtive-K0}
\end{align}
where the tensor $\langle \vc{r}\Ja+\Ja\vc{r} \rangleb$
%$\ku\tAa = -\intd{\Reals^3~}{}{ \vc{r}\vc{r}\gradr\cdot(\Ja\psib)}{\pt{r}}$ 
is the only new term compared to \cite[p. 44]{Larson.1988.1}. 
%The use of $\ku$ as a timescale will be clarified
%in \refsec{sec-motors} and is only a formal convenience,
%since the timescale that will arise from active terms is in general not independently accessible.
From \eqref{eq-contractilitytensor}, we have $\langle \vc{r}\Ja+\Ja\vc{r} \rangleb = (\ku+\ks)\tAa -\ks \tA$.

We denote the upper-convected derivative of a tensor $\ts{T}$ by 
\begin{align}
\ucm{T} = \ddt{\ts{T}} -\tsL\cdot\ts{T}-\ts{T}\cdot\tsL\,\trsp
\label{eq-uc}
\end{align}
The fact that this particular objective derivative arises is due
to the contra-variant nature of the vector $\vc{R}$ which represents the fibrous microstructure 
\cite{Hinch-Harlen.2021.1}. It is thus not an arbitrary choice but an intrinsic property of materials
formed of a network of entropic chains. Note also that the nonlinearity in \eq{eq-uc} is not likely
to be eliminated by an order of magnitude analysis: indeed, if the shear rate is of order $1/T$,
then both the partial derivative in time and the velocity gradient $\tsL\,\trsp$ are
of order $1/T$. The legitimate linearisation of such a viscoelastic
constitutive equation is thus a purely viscous constitutive equation.

Since we have chosen to assume
constant rates of binding and unbinding, $\mathcal{K}_0 = \psib - \au(\kb/\ku) \psi_0$, we
obtain \JEadds{that 
$$
-\ku \intd{\Reals^3}{}{\mathcal{K}_0(\psib) \vc{r}\vc{r}}{\pt{r}} = -\ku \tA + \au\kb\beta^{-2}\ts{I}.
$$
Altogether, this leads to:}
\begin{align}
\tau\ucm{A} &= \tA_0 - \tA + \tAa.
\label{eq-relaxtive}
\end{align}
where we have identified $\tA_0 \JEadds{= \au\kb/(\ku+\ks) \left<\vc{r}\vc{r}\right>_0}$ as the
long-time limit of $\tA$ in the absence of flow and activity, thus corresponding to relaxed chains.
Comparing with \refsec{sec-stress} and since $\langle \vc{r}\vc{r}\rangle_0 = \frac{\beta^{-2}}{2}\ts{I}$, we set $a_0=\au\kb/(2\ku+2\ks)\simeq \frac{1}{2}$ for $\ks\ll\ku\ll\kb$.
\JEadds{We have defined the \emph{relaxation time} of the network as $\tau = (\ku+\ks)^{-1}$,
thus based on the average rate of the unbinding-rebinding process and of the internal slippage
process.
}

We can also rewrite the specific free energy as the sum of the entropic energy of bound and unbound chains,
$$
\varphi_{\mathrm{e}} = \frac{\kappa}{2}\intd{\Reals^3}{}{(\psib+\psiu)\pt{r}^2}{\pt{r}} 
	= \frac{\kappa}{2}\tr\tA + \frac{3}{2}\au k_BT.
$$

Using the definition of $\tsigma$, and observing that $ \ucm{I} = -2\tsD $, we find:
\begin{align}
\tau\ucm{\sigma} + \tsigma = 2\tau G \tsD + \tsiga
\label{eq-veac}
\end{align}
in the case when $\mu=0$, %\mu_b=0$,
which is the upper-convected Maxwell constitutive equation with relaxation time $\tau$
and short-time elastic modulus $G = \JEadds{\frac{\kb}{\ku}\au\nu k_BT \simeq} \nu k_BT$ for $\ku\ll\kb$. 
The Oldroyd-B model can be obtained with a nonzero liquid bath viscosity $\mu$.
There is a single relaxation time that appears in this model, which combines two relaxation processes of different nature (unbinding/rebinding $\ku$ and internal slippage $\ks$). One could consider extensions to
multiple relaxation modes in the spirit of the Lodge model \cite{Lodge.1956.1,Bird+.1987.2}
or other models involving multiple crosslinks per chain \cite{Broedersz+MacKintosh.2010.1}.
If $\Ja$ is nonzero, there is an additional term, $\tsiga = \nu\kappa \tAa$ 
that can be interpreted as a dynamic
prestress \cite{Erlich+Wyatt.2022.1} and can be identified with the active stress of the active gel theory \cite{Kruse+.2005.1}.

%Note that the Maxwell equation also corresponds to anisotropic elasticity:
%linearising the material response at short times around an anisotropic
%state characterised by a microstucture $\tA_1$, 
%we find that the apparent elastic modulus is $G\ts{I}+\nu\kappa(\tA_1-\tA_0)$.
%\PR{Je ne comprends pas la phrase précédente. Tu es sur qu'elle est nécessaire ? C'est bizarre ce module tensoriel. Je comprendrais si c'était un tenseur d'ordre superieur à la rigueur mais là je ne saisis pas.}

\subsection{Dissipation and viscoelastic relaxation}
\label{sec-relax}

Using the definition of $\tA$ and $\tA_0$, we evaluate the part of the dissipation which is due to the unbinding--rebinding dynamics:
%With the above constitutive choices for $p$ and $\tsigma$, we now have:
%the dissipation:
\begin{align*}
%\mathcal{D} &= \mathcal{D}_w + \mathcal{D}_r + \mathcal{D}_a %-\ddt{\mathcal{F}_{\!\mathrm{a}}}
%	\\
%	&= \mathcal{D}_w 
\mathcal{D}_{\mathrm{r}} = 
	 \frac{1}{2}\intd{\Omega(t)}{}{\ku \nu \kappa (\tA-\tA_0):\ts{I} }{\pt{x}}
	%- \intd{\Omega(t)}{}{\left(
	%	\nu\kappa\langle\Ja\vc{r}\rangleb  + \nu \ddt{\varphi_{\mathrm{a}}}  %\right)}{\pt{x}}.
\end{align*}
The term $-\frac{\nu\kappa}{2}\tA_0:\ts{I} = -\frac{3}{2}\nu k_BT$ for $\ks\ll\ku\ll\kb$, corresponds to the
thermal energy due to the equipartition of the unbound chains with the bath.
It is thus a lower bound for the
elastic energy $\nu\kappa\tA$. % for all conditions at thermal equilibrium,
%\JE[Indeed it is shown that if the initial condition for $\tA$ is such that the 
%dumbbell beads have a thermal agitation at least equal to the one of the liquid bath,
%$\det\tA(t=0) \geq \det\tA_0$,
%then $(\tA-\tA_0):\ts{I}\geq 0$ and thus $\mathcal{D}_r\geq 0$ for all times 
See \cite{Hu-Lelievre.2007.1} and \refapp{app-trace} for a proof that this is the case
with $\tA$ obeying \eqref{eq-relaxtive} as long as the initial condition is admissible, $\det(\tA(t=0))\geq\det(\tA_0)$, and only if the active stress is contractile, that is $\tAa$ positive semi-definite.

\JEadds{
Note that as seen in \refsec{sec-motors}, the active crosslinkers are thermodynamically
required to have a `slippage' rate which has a dissipative trace of the form $\frac{1}{2}\ks\nu\kappa \ts{A}:\ts{I}$ 
similar to the relaxation via unbinding, $\frac{1}{2}\ku\nu\kappa \ts{A}:\ts{I}$. This combination gives the final relaxation time of the network $\tau=(\ku+\ks)^{-1}$ in \eqref{eq-relaxtive}.
}

We can make use of the relaxation  \eq{eq-relaxtive} to justify that $\mathcal{D}_{\mathrm{r}}$ corresponds to viscoelastic relaxation:
$$
\mathcal{D}_{\mathrm{r}} 
 = \frac{1}{2}\intd{\Omega(t)}{}{ \nu\kappa\ku (-\tau\ucm{A}+\tAa):\ts{I} }{\pt{x}} 
$$
Thus, in the passive viscoelastic case $\tAa=\ts{0}$, 
the relaxation term is indeed proportional to the (negative) trace of $\ucm{A}$ 
corresponding to the microstructure dissipating elastic energy.
In the active case where $\tAa\not=\ts{0}$ and is positive semi-definite, we still have $\mathcal{D}_{\mathrm{r}} \geq 0$, however the active
strain is superimposed to the relaxation dynamics.

\subsection{Summary}
\label{sec-summary}

Here we take the limit $\ks\ll\ku\ll\kb$, and hence $\tau=1/\ku$, and summarise the equations to obtain a closed model. The flow rate has to be such that $|\grad\vc{v}|\lesssim \kb$.
The total specific free energy is then
\begin{align}
\varphi = \varphi_\mathrm{e}+\varphi_\mathrm{a} 
	&= 
	    \frac{\kappa}{2}\tr\tA(\pt{x},t)
		-\Deltamu \langle\xic\rangleb(\pt{x},t),
\label{eq-freeenergy}
\end{align}
where $\kappa=2k_B T \beta^2$. The thermodynamics of the system constrains 
the time evolution of the microstructure tensor $\tA$, which
%and in some measure of the advancement $\xic$ of the ATP hydrolysis by molecular motors.
%\PR{Je ne mettrais pas la fin de la phrase, je pense qu'elle est dure à comprendre pour qqn qui ne connait pas bien les gels actifs. On ne regarde pas $\xi_a$ de toute façon.}
%
%The microstructure tensor $\tA$ 
relaxes %through the unbinding dynamics
towards a state that can be prestrained by molecular motors with the dynamics given
in \eq{eq-relaxtive},
\begin{align*}
\tau\ucm{A} &= -\tA + \tA_{0} + \tAa,
\end{align*}
where the microstructure passive equilibrium tensor is $\tA_0 = \frac{\beta^{-2}}{2} \ts{I}$. 
The contractility is 
$\tAa=a_{\mathrm{a}}^2\langle\ts{Q}\rangleb$, where $a_{\mathrm{a}}^2={2 \ell_{\mathrm{a}}\beta \Deltamu}/{\kappa}$ 
characterises the motor activity.
The orientation tensor
$\langle\ts{Q}\rangleb = \langle \vc{r}\vc{r}/\pt{r}^2 \rangleb$ is not \JEadds{equal} to the microstructure tensor $\tA=\langle \vc{r}\vc{r} \rangleb$.
It can in some cases be deduced from the symmetries of the flow,
see \refsec{sec-examples} and e.g.\ \citetext{Dicko+Etienne.2017.1}. It is also sometimes assumed to
be isotropic. It could also be approximated to $\tA/(\tA:\ts{I})$, or finally could be
calculated using a multiscale model that would solve \eq{eq-smolu} explicitely 
\cite{Jourdain+LeBris.2004.1}.

The density of chains evolves with
\begin{align}
\dedt{\nu} + \div(\nu \vc{v}) 	&= 0. 
\label{eq-nu}
\end{align}
The total stress tensor is $-p\ts{I}+\tsigma$, where the extra stress
$\tsigma$ originates from the network configuration and from possible viscous contributions,
\begin{subequations}
\label{eq-constitutive}
\begin{align}
\tsigma&=\nu\kappa(\tA-\tA_{0}) + 2\mu\tsD, %+ \mu_b(\div\vc{v})\ts{I}, 
\\
\intertext{
and %\JE[in the incompressible case]{delete}
the pressure $p$ is the Lagrange multiplier
that ensures 
}
\div \vc{v} &= 0.
\end{align}
\end{subequations}
Initial conditions are required for $\nu$ and $\tA$. The momentum balance and associated
boundary conditions are given in \eq{eq-momentum}, which allows to solve for the velocity $\vc{v}$.

The present model \eqs{eq-relaxtive},\eqref{eq-constitutive},\eqref{eq-momentum},\eqref{eq-nu}
can thus be solved for the time evolution of the 
microstructure tensor $\tA$ and extra stress tensor $\tsigma$ and pressure $p$, network
velocity $\vc{v}$ and density $\nu$. Note that \eqs{eq-relaxtive},\eqref{eq-constitutive} can be combined to eliminate the microstructure tensor $\tA$ and solve directly in terms of the extra stress $\tsigma$, giving \eq{eq-veac}.

When the characteristic time of the flow is very long compared to the relaxation time $\tau$, a viscous limit of \eq{eq-veac} eliminating $\tau\ucm{\sigma}$ but retaining the anisotropic active stress $\tsiga$ can be taken, this is the model used e.g. in \cite{Dicko+Etienne.2017.1}. Otherwise, the nonlinear objective derivative $\ucm{\sigma} = \lddt{\tsigma}-(\grad\vc{v})\trsp\tsigma-\tsigma\grad\vc{v}$ is required in order to account for the entropic nature of the microstructure. This is the case even when the problem can be reduced to a one-dimensional case ($\tsigma=\sigma_{xx} \vc{e}_x\vc{e}_x,\,\vc{v}=v_x\vc{e}_x$) \cite{Etienne+Asnacios.2015.1,Roux+Etienne.2016.1} since, contrarily to the co-rotational objective derivative \cite{Recho-Truskinovsky.2013.1}, a nonlinear coupling remains in the longitudinal component $(\ucm{\sigma})_{xx} = \lddt{\sigma_{xx}}-2(\partial_x v_x)\sigma_{xx}$. Taking values from e.g.\ recoil after laser ablation of actomyosin experiments \cite{Saha+Grill.2016.1}, one can estimate that the order of magnitude of this nonlinear term is similar or larger than the viscous one. This could provide an alternative way to test experimentally whether a biopolymer network exhibits entropic or enthalpic elasticity.

\section{Multiplicative strain decomposition framework}
\label{sec-multiplicative}

Multiplicative decomposition of the deformation gradient is commonly used 
for thermoelasticity and elastoplasticity applications \cite{Lubarda.2004.1}. 
It has also proven very useful in biomechanics, where the main applications 
have been the understanding of residual stress originating from growth in
soft tissue \cite{Rodriguez+McCulloch.1994.1,Taber.1995.1} but also plants or
hard tissue \cite{Goriely.2017.1}.
In these contexts, growth is then considered as a prestrain.
Prestrain can also be used to model contractility, as e.g.\ in \citetext{Fierling+Etienne+Rauzi.2022.1},
and indeed there exist formal analogies \cite{Erlich+Wyatt.2022.1}.

For liquid-like systems, an additional phenomenon is the microstructure relaxation. Many numerical models that
aim at reproducing the phenomenology of microstructure relaxation use an algorithm that can formally be likened to
morphoelasticity: at each time step, solve for the elastic deformation 
relative to some intermediate configuration,
this configuration being the equilibrium configuration of the previous time step. Formally, this corresponds to a
morphoelastic model where the anelastic defomations are the viscous-like deformations cumulated through time.
This is extremely convenient for e.g.\ the dynamics of slender liquid visco-elastic structures
\cite{Bergou-Audoly+.2010.1,NestorBergmann+Sanson.2022.1}. 
However this approach is very crude in the sense that it cannot describe any dynamics at a characteristic time close
or smaller than the relaxation time of the material, and of course that its thermodynamics are uncontrolled.
Recently, a multiplicative strain decomposition has been introduced \cite{Alrashdi-Giusteri.2024.1} for viscoelatic liquid models.

Here, we make use of the multiplicative decomposition of the deformation gradient formalism
and derive the evolution equation of the anelastic part of the deformation so that it matches
the active viscoelastic model developed in the previous section. 

\begin{figure}
\centering\includegraphics[width=.5\textwidth]{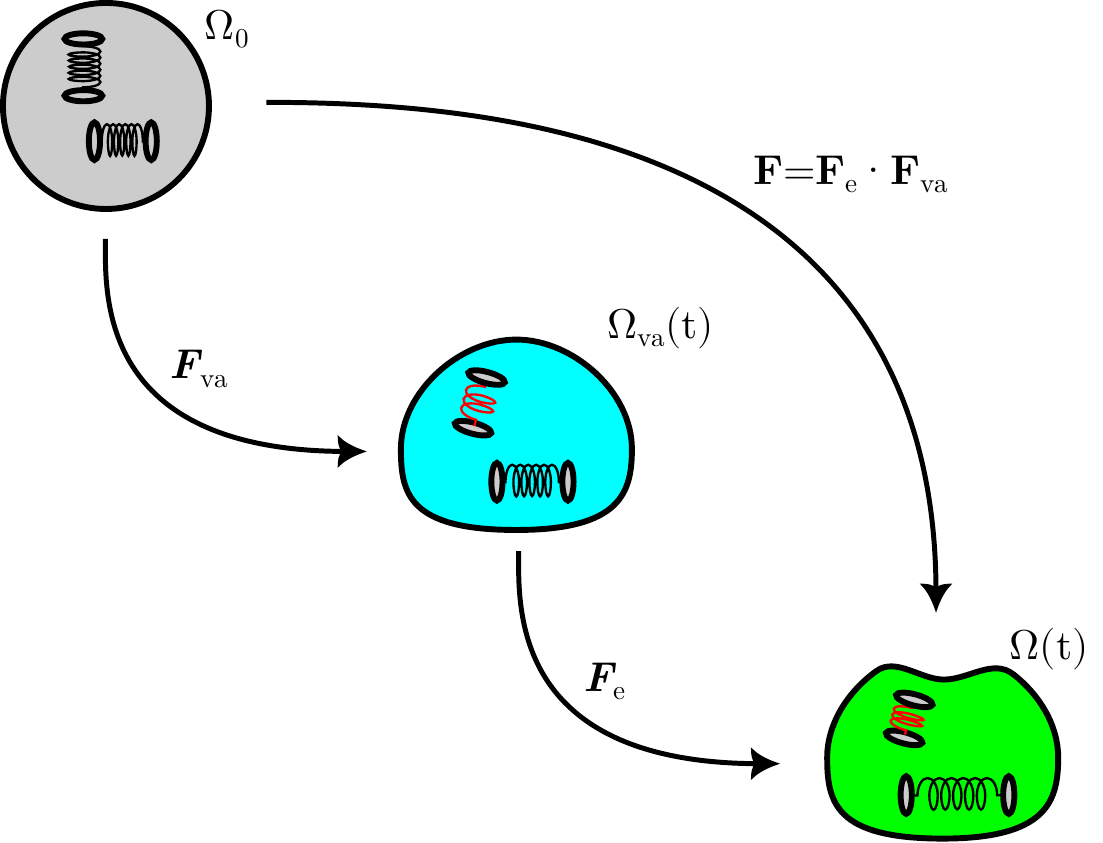}
\caption{Multiplicative decomposition of the deformation gradient for viscoelastic active fluids.}
\label{fig-decomp}
\end{figure}

We choose the decomposition as $\ts{F}=\Fe\cdot\Fva$, illustrated in \fig{fig-decomp}, where 
$\Fe$ corresponds to elastic deformations of the microstructure,
and $\Fva$ corresponds to both viscous-like deformations due to the relaxation of the elastic microstructure and active deformation due to a chemically-driven growth or contraction of the microstructure. Our objective is to obtain a set of 
equations equivalent to the model of \refsec{sec-summary} in terms of $(\Fva,\Fe,p)$.

%\JE[We now aim at writing the free energy
%in terms of the active deformation tensor in addition to the elastic
%strain, $\varphi_{\text{mult}}(\Fe,\Fva) = \varphi(\tA,\xic)$
%as defined in \eq{eq-freeenergy}.
%Thus $\Fe$ should be related to $\tA$ and $\Fva$ with $\xic$.]{No we don't do that...}

The tensor $\tA$ is by construction symmetric and positive semidefinite (and
positive definite as long as $\mathrm{Supp}\,\psib$ has nonzero measure, 
which can be ensured by smooth initial conditions and the affine or diffusive nature of the fluxes in \eq{eq-smolu}). 
Thus Cholesky factorisation guarantees the existence of an upper triangular tensor $\ts{U}$ such that 
$\tA = \ts{U}\trsp\cdot\ts{U}$. % = \ts{U}\trsp\cdot\ts{O}\trsp\cdot\ts{O}\cdot\ts{U}$ with $\ts{O}$ an orthogonal matrix. 
Thus we can define $\Fe$ so that $\tA$ is either $\alpha\Fe\trsp\cdot\Fe$ or $\alpha\Fe\cdot\Fe\trsp$, with $\alpha$ a constant,
imposing that the upper diagonal matrix
in the QR-decomposition of $\Fe$ is $\ts{U}$. Since in both choices, the invariants $\Fe\trsp\cdot\Fe:\ts{I}=\Fe\cdot\Fe\trsp:\ts{I}=\Fe:\Fe$ are 
identical, we make our choice in order to be able to define the corresponding $\Fva$ conveniently. As in \cite{Califano-Ciambella.2023.1}, we remark that
the (Eulerian) \emph{left} Cauchy--Green tensor $\ts{B} = \ts{F}\cdot\ts{F}\trsp$
is such that $\dts{B} = \tsL\cdot\ts{B} + \ts{B}\cdot\tsL\,\trsp$ and thus, that its upper convected derivative
is zero, which guides us to choose $\Fe$ such that $\Fe\cdot\Fe\trsp=\alpha\tA$.

As a result, $\tA = \JEadds{\alpha^{-1}} \ts{F} \cdot\Cva^{-1} \cdot\ts{F}\trsp$ with $\Cva^{-1} = \Fva^{-1}\cdot\Fva\mtrsp$, and by derivation
$$
\ucm{A} = \dot\tA - \tsL\cdot\tA - \tA\cdot\tsL\trsp = \JEadds{\alpha^{-1}} \ts{F} \cdot\ddt{\Cva^{-1}}\cdot \ts{F}\trsp
$$
which allows to use \eq{eq-relaxtive} to set the dynamics of $\Cva^{-1}$,
$$
\tau \ts{F} \cdot\ddt{\Cva^{-1}} \cdot\ts{F}\trsp = -\ts{F} \cdot\Cva^{-1}\cdot \ts{F}\trsp + \JEadds{\alpha}\tA_0 + \JEadds{\alpha}\tAa
 $$
which yields, since $\tA_0=a_0\beta^{-2}\ts{I}$,
\begin{align}
\tau \ddt{\Cva^{-1}} = -\Cva^{-1} + \alpha  a_0 \ts{C}^{-1} +  \JEadds{\alpha}\ts{F}^{-1}\cdot\tAa\cdot\ts{F}\mtrsp.
\end{align}
\JEadds{where $\alpha$ remains to be determined.}
In the permanent regime and in the absence of activity $\tAa$, elastic strains relax and thus $\Cva^{-1}$ tends to $\ts{C}^{-1}$. We
thus set $\alpha = \beta^2 a_0^{-1}$ using this limit behaviour. \JEadds{For $\kb \gg \ku$, this is thus $\alpha = 2\beta^2$. For convenience, we define the \emph{active strain tensor} $\tEa = -\alpha\tAa$. Since our theory requires that $\tAa$ is positive semi-definite (\refapp{app-trace}), $\tEa$ is negative.}
%\JE{Qu'est ce que tu penses de la terminologie...? et du signe... j'ai pris la convention avec $\tEa = - \alpha\tAa$ négatif pour être cohérent avec un prestrain contractile.}
\JEadds{Thus, the active stress $\tsiga$ defined in \refsec{sec-veac} is $\tsiga = -G\tEa$. Note that our model of active crosslinkers leads to nonpositive eigenvalues of $\tEa$ (interpreted as a contractile, thus negative, prestrain) and hence nonnegative ones of $\tsiga$.}

In summary, the model of \refsec{sec-summary} can now be rewritten using the multiplicative decomposition.
Given $\tEa,G,\mu,\tau$, find $(\Cva,\JEadds{\vc{u}},p)$ such that:
\begin{subequations}
\begin{align}
    &\tau \ddt{\Cva^{-1}} 
        = -\Cva^{-1}  
        +  \ts{F}^{-1}\cdot\JEadds{(\ts{I} - \tEa)}\cdot\ts{F}\mtrsp,
\label{eq-morpho-va}
\\
    &\grad\cdot\tsigma[\Cva^{-1},\vc{u}] - \grad p = \vc{0},
\\
    &\det \ts{F}[\vc{u}] = 1,
%\end{align}
%\end{subequations}
\\
%    \Fe\cdot\Fe\trsp &= \Be
%\\
%    \ts{F}\trsp\cdot\ts{F} &= %\Fe\trsp\cdot\Cva\cdot\Fe,
%\\
\intertext{where}
%\begin{align*}
    \tsigma &= G( \Be[\vc{u}] - \ts{I} ) + \mu(\tsL+\tsL\,\trsp),
    \label{eq-morpho-constit}
    \\
    \ts{F} &= \ts{I}+\grad\vc{u}\trsp,
    \qquad
    \Be = \ts{F}\cdot\Cva^{-1}\cdot\ts{F}\trsp,
    \qquad
    \tsL = \dts{F}\cdot\ts{F}^{-1},
\end{align}
\end{subequations}
subject to an initial condition on $\Cva$ and boundary conditions on $(\tsigma - p\ts{I})\cdot\vc{n}$.
\JEadds{%The boundary conditions will in general allow to set $\ts{F}$ which is here defined up to a solid rotation, unless they are all of Neumann type.
As discussed above, if $\Fva$ and $\Fe$ are needed, then it has to be noted that any local rotation $\ts{O}(\pt{x})$ of the intermediate configuration is permitted, 
since for $\ts{O}$ an orthogonal matrix, $\Fe'=\Fe\cdot\ts{O}$ and $\Fva'=\ts{O}\trsp\cdot\Fva$ are
also solutions of the problem with unchanged $\ts{F},\Be,\Cva$. %\JE[This can be understood as a consequence of the material indifference
%principle \cite{Joseph}.]{Ptet un peu glissant... }
To fix this, one can e.g.\ impose that $\Fe\trsp$ is upper-triangular with
positive diagonal coefficients.}

\section{Example applications}
\label{sec-examples}

We present here two examples in which the dynamics of an actively contractile structure can be solved in a straightforward manner thanks to the multiplicative decomposition of the deformation gradient. 
Anisotropic geometry and anisotropic contractility is present in both examples, either in aligned or orthogonal configurations. 
In both examples we neglect the liquid bath viscosity $\mu$ and take the orientation tensor $\langle \ts{Q} \rangleb$ as a given input. 

\subsection{Actin stress fibres: an active viscoelastic beam in an elastic \JEadds{environment}}
\label{sec-beam}

\begin{figure}[btp]
\textbf{A$\!\!\!\!$}\includegraphics[width=\textwidth]{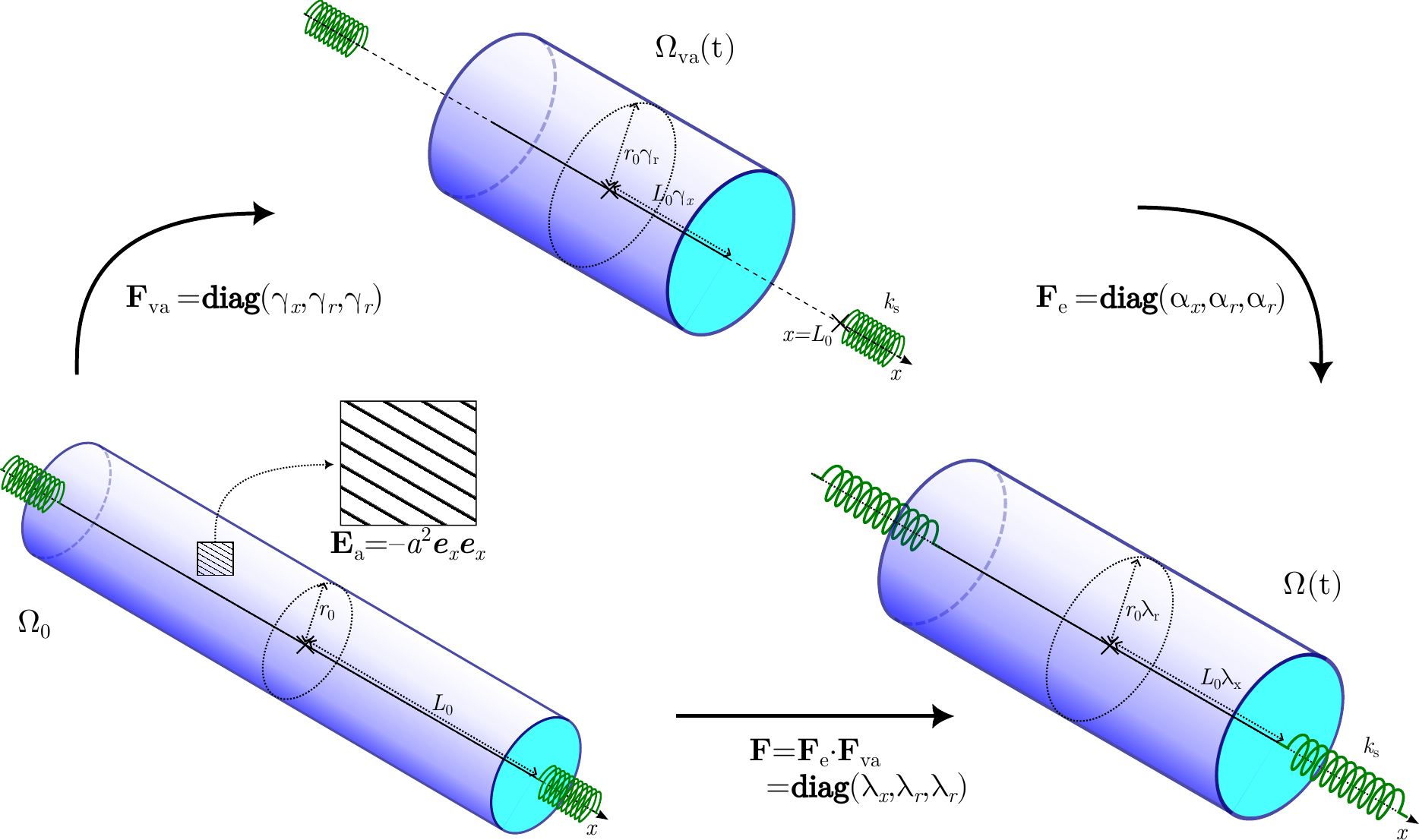}\\
\textbf{B}\includegraphics[width=.30\textwidth]{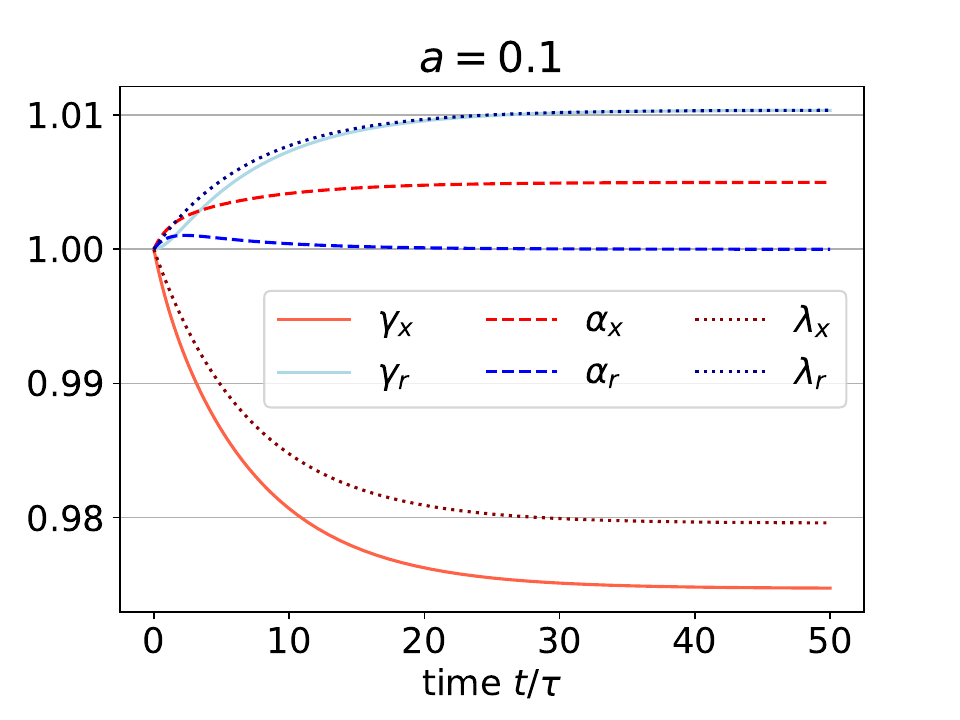}
\textbf{C}\includegraphics[width=.30\textwidth]{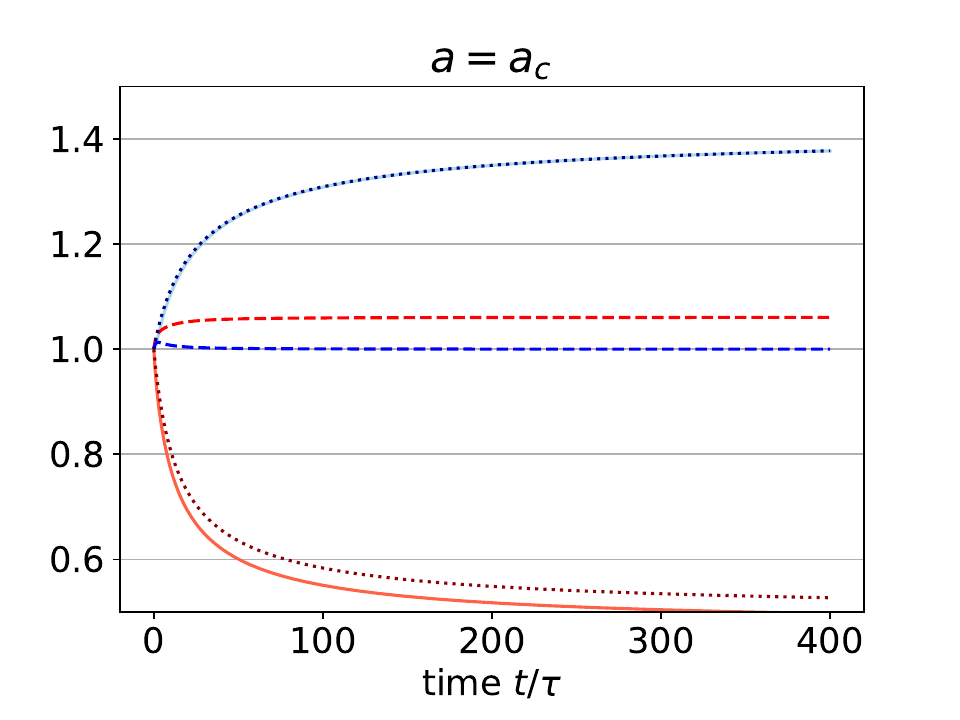}
\textbf{D}\includegraphics[width=.30\textwidth]{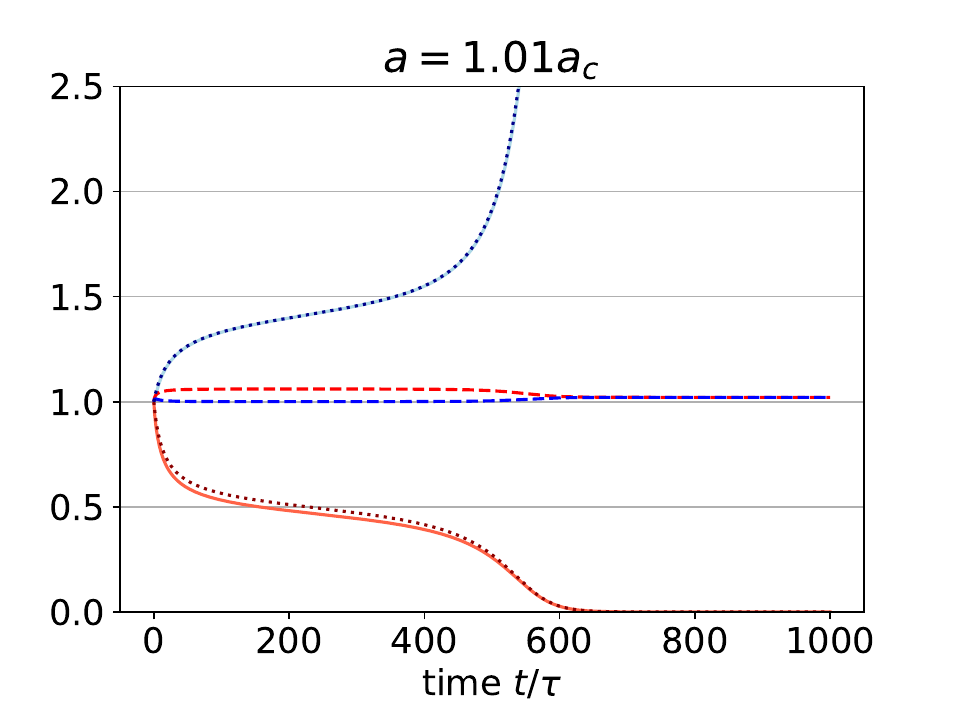}
\caption{
Uniaxially contractile beam interacting with an elastic environment.
\textbf{A}, The initial configuration $\Omega_0$ of the beam, before the active strain $\tEa$ has been applied,
is chosen here at equilibrium with the external springs.
The visco-active deformation $\Fva(t)$ gives the intermediate virtual configuration $\Omega_{\mathrm{va}}(t)$,
a configuration which is not compatible with the external forces and does not obey incompressibility.
The elastic deformation $\Fe(t)$ restores both of these requirements in the current configuration $\Omega(t)$.
The time evolution of $\Fva(t)$ is prescribed by the relaxation dynamics of the microstructure $\tA$
towards a state where the elastic stress is equal to the active stress, 
$\tsigma(t_{\mathrm{eq}}) = \nu\kappa(\tA(t_{\mathrm{eq}})-\tA_{0}) = -G\tEa$.
\JEadds{
\textbf{B-D}, Dynamics for a contractile strain $\tEa=-a^2 \vc{e}_x\vc{e}_x$
for different values of $a$ and $G/E_s=1$. In all cases, the contractile
strain drives an initial decreasing $\gamma_x$ (active shortening) which is partly balanced by an increasing $\alpha_x$ (elastic stretch), resulting in a lesser shortening of the current configuration $\lambda_x$. Incompressibility imposes a swelling in the radial direction, $\lambda_r>1$, which after a transient elastic stretch $\alpha_r>1$ drives a viscoelastic relaxation towards $\gamma_r=\lambda_r$.
\textbf{B}, for $a=0.1$, the contraction is 90\% of the final contraction at $t\approx 17\tau$. 
\textbf{C}, for $a$ equal to the critical value $a_c\approx 0.35$, the contraction is 90\% of the final contraction at $t\approx 195\tau$. 
\textbf{D}, for a contractility slightly above the critical value, the contraction reaches the critical value $0.5$ at $t\approx 229\tau$, the beam
then collapses tending to $0$ length while the radial stretch diverges.
}}
\label{fig-beam}
\end{figure}

We model a slender active viscoelastic beam, which can for example represent a stress fibre in a cell 
with adhesion at its ends only \cite{Katoh+Fujiwara.1998.1}.
%\JE[viscosities $\mu=\mu_b=0$.]{remove bulk}
It initially spans the distance from $(-L_0,0,0)\trsp$ to $(L_0,0,0)\trsp$ and has radius $r_0$.

We assume that the chains within the beam are all oriented along the main axis $x$ of the beam, leading
to $\langle\ts{Q}\rangleb = \vc{e}_x\vc{e}_x$ and $\tEa = -a^2 \vc{e}_x\vc{e}_x$. 
The external forces are supposed to be zero along the beam itself (no friction), except for
tractions $\pm F\vc{e}_x = \pm k_s L_0(1-\lambda_x)\vc{e}_x$ applied at each end $x=\pm L_0$ of the
beam and which correspond to the elastic resistance to the deformation of the environment which behaves
as a spring of stiffness $k_s$. We define $E_s = k_s L_0/(\pi r_0^2)$.

With this geometry and restricting load to the longitudinal direction only, \JEadds{the deformation 
gradient and stress tensors are diagonal tensors,
$\ts{F}=\diag(\lambda_x,\lambda_r,\lambda_r)$
and $G( \Fe\cdot\Fe\trsp - \ts{I} ) = \diag(\sigma_{xx}, \sigma_{rr}, \sigma_{rr})$.
Thus $\Fe\cdot\Fe\trsp$ is diagonal, and we can choose}
$\Fe=\diag(\alpha_x,\alpha_r,\alpha_r)$,
$\Fva=\diag(\gamma_x,\gamma_r,\gamma_r)$. By definition, $\lambda_i = \alpha_i\gamma_i$, $i\in \{r,x\}$.

%\JE[Taking the pressure to be a Lagrange multiplier that imposes incompressibility,]{remove} 
We have the
equilibrium equations:
\begin{align}
G(\alpha_x^2 - 1) - p &= E_s(1-\lambda_x)\lambda_r^{-2}
\\
G(\alpha_r^2 - 1) - p &= 0
\\
\lambda_r^2 \lambda_x &= 1
\end{align}
Thus, subtracting the $r$ equation from the $x$ one to eliminate pressure, and using
the incompressibility condition,
$$
\lambda_x^2 - \lambda_x + \frac{G}{E_s}(\alpha_x^2-\alpha_r^2) = 0.
$$
From \eq{eq-morpho-va}, we have:
\begin{align}
2\tau \dot{\gamma}_x &= (\alpha_x^{2}-1-a^2)\alpha_x^{-2}\gamma_x
\\
2\tau \dot{\gamma}_r &= (\alpha_r^{2}-1)\alpha_r^{-2}\gamma_r
\end{align}
For $a\leq a_c = \sqrt{E_s/(4G)}$, there is an
admissible long times solution with $\alpha_x^2=1+a^2$, $\alpha_r=1$,
for which $\gamma_i>0$ and leading to $\lambda_x= \frac{1}{2}(1+\sqrt{1-4Ga^2/E_s})$.
See \fig{fig-beam}\textbf{B,C} for the dynamics leading to that asymptotic state.
For $a>a_c$, the beam reaches length $L_0/2$ in finite time, which (provided that $\langle\ts{Q}\rangleb$ is unchanged) leads to a catastrophic collapse: indeed, the broadening section
of the beam then dominates over its shortening, and the external traction (per unit surface of the section) then decreases with increasing deformation.
The external traction is thus unable to balance the internal contractile stress for any deformation and the material flows towards a zero length of the "beam".

The relevance of this model for actomyosin-based systems is
discussed in \cite{Etienne+Asnacios.2015.1}, where we also find that a viscoelastic liquid material
with an internal active stress  adapts in length to the external stiffness $E_s$. It may also be seen as a modelling framework for the so-called actin ventral stress fibres \cite{Deguchi+Sato.2006.1,Fage+Asnacios.2024.1}.

\begin{figure}[btp]
\textbf{A$\!\!\!\!$}\includegraphics[width=0.64\textwidth]{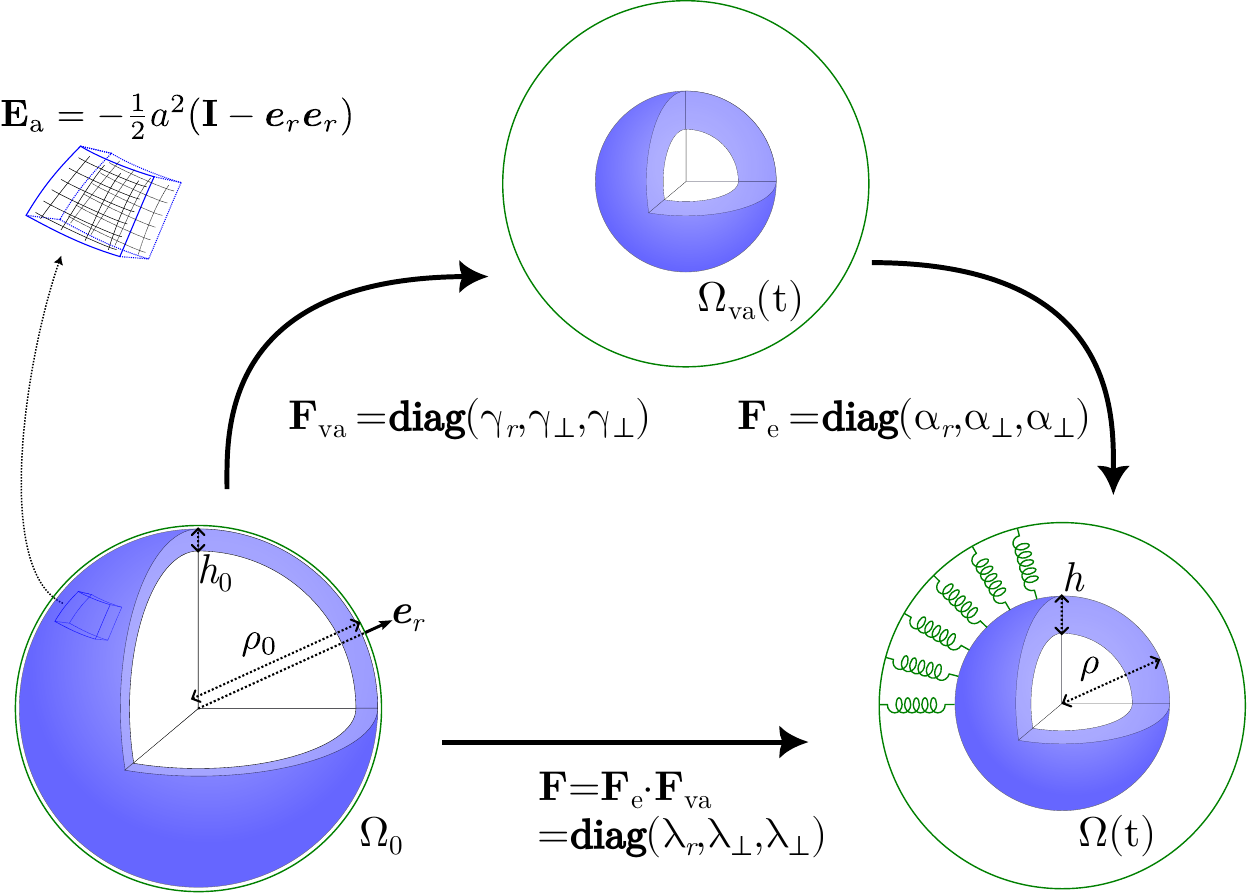}
\begin{minipage}[b]{0.34\textwidth}
\textbf{B}\includegraphics[width=0.95\textwidth]{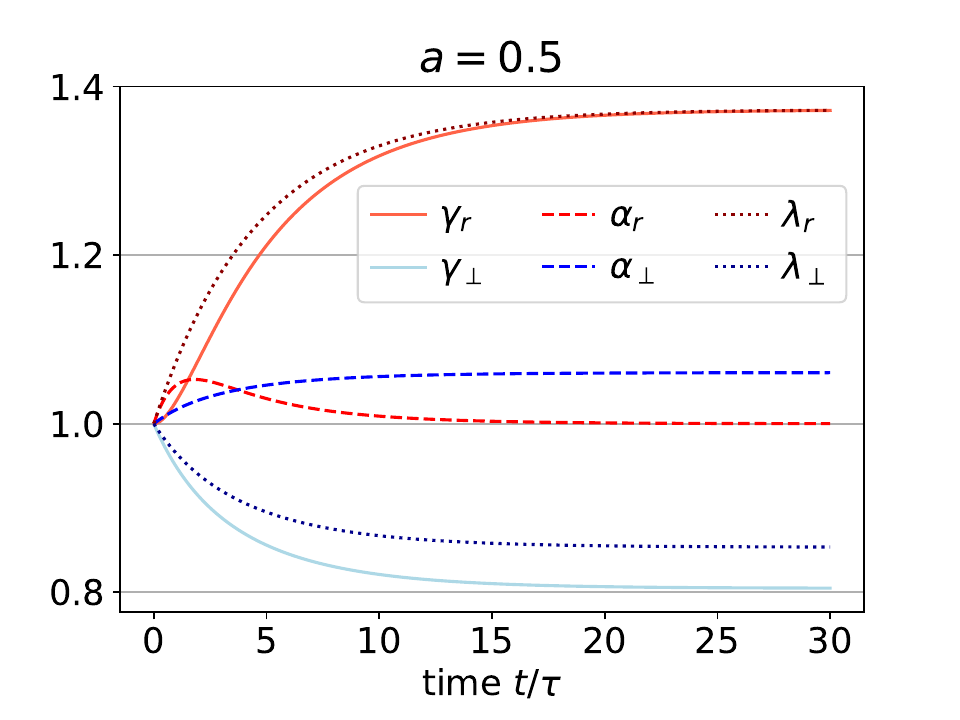}\\
\textbf{C}\includegraphics[width=0.95\textwidth]{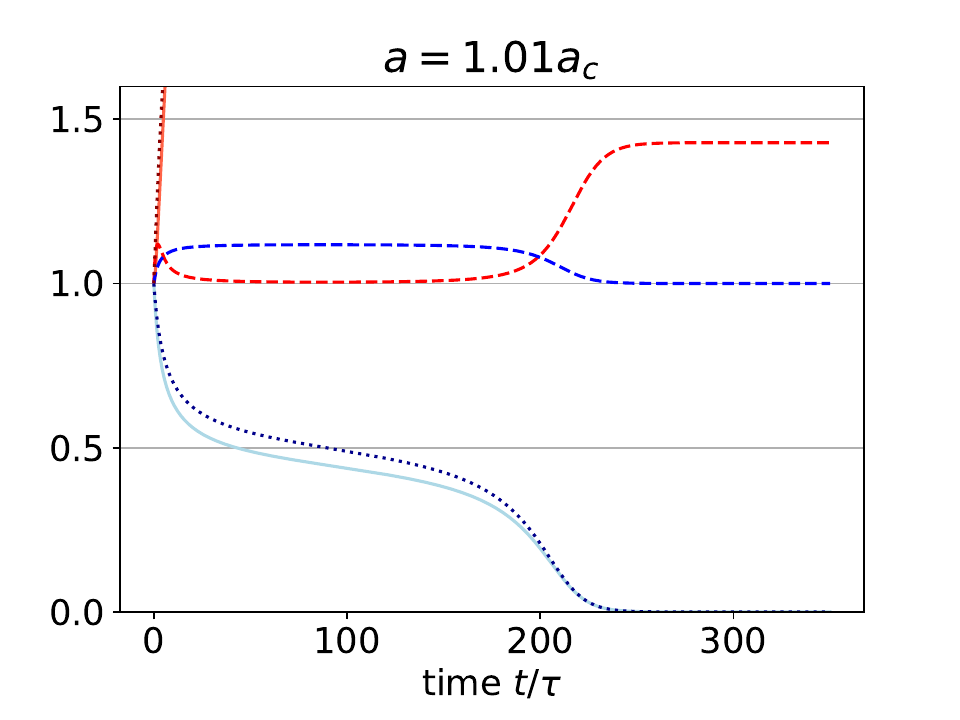}
\end{minipage}
\caption{Tangentially contractile sphere bound elastically to a fixed sphere. \textbf{A}, The initial, intermediate and current configurations of the contractile sphere. In the current configuration, mechanical balance with the Winkler foundation which binds it to a fixed sphere of radius $\rho_0$ has to be verified. \textbf{B,C}, Dynamics for $a_c = \sqrt{2}/2$ and an orthoradial contractile strain $\tEa=-\frac{1}{2}a^2(\ts{I}-\vc{e}_r\vc{e}_r)$ with two different magnitudes of $a$, respectively below and above the critical value $a_c$. Transient behaviours similar to the case of the contractile beam are observed, although the axes in which they appear differs due to the different geometry and active stress orientation.
}
\label{fig-cyst}
\end{figure}

\subsection{Apical luminal surface: actively contractile spherical shell}

Next, we turn to a 3D case where contractility is tangential to the surface of the structure. This can be inspired e.g. by the apical surfaces of cells organised as a spherical monolayer around a central lumen in a cyst \cite{O'Brien+Mostov.2002.1}, which can be represented by a thin, closed surface in elastic interaction with a spherically shaped basement membrane that contains it. For simplicity we will treat this surface as perfectly permeable, in reality the water flux itself is under an active control. 

Figure \ref{fig-cyst} illustrates this spherical shell geometry, with initial outer radius $\rho(t=0) = \rho_0$ and finite thickness $h(t=0) = h_0\ll \rho_0$. We assume that it contracts tangentially in an isotropic manner and consider only solutions retaining spherical symmetry. Their stability will be studied elsewhere. Thus, $\tEa = -\frac{1}{2} a^2 (\ts{I} - \vc{e}_r\vc{e}_r)$. 
The elastic interaction with a fixed outer sphere of radius $\rho_0$ is modelled by a Winkler elastic foundation characterised by a spring stiffness $k_s$ and a density $\Sigma_0$ on the outer sphere, and thus $\Sigma_s = \Sigma_0 (\rho_0/\rho)^2$ on the outer surface of the shell.
The boundary conditions are thus,
$$
\ts{\tau}(\rho)\cdot\vc{e}_r 
= f_\text{ext} \vc{e}_r 
= E_s\frac{\rho_0(\rho_0-\rho)}{\rho^2}\vc{e}_r,
\qquad
\ts{\tau}(\rho-h)\cdot\vc{e}_r = \vc{0},
$$
with $E_s = \Sigma_0 \rho_0 k_s$. Spherical symmetry imposes that $\ts{\tau}$ and $\ts{F} = \diag(\lambda_r,\lambda_\perp,\lambda_\perp)$ are diagonal tensors.
Hence, the current configuration is a spherical shell of outer surface area $4\pi(\lambda_\perp\rho_0)^2$, and thus radius $\rho=\rho_0\lambda_\perp$ and thickness $h=h_0\lambda_r = h_0/\lambda_\perp^2$.
From \eqref{eq-morpho-constit} and $\Fe$ being chosen upper diagonal, it must have the form $\Fe = \diag(\alpha_r,\alpha_\perp,\alpha_\perp)$ and hence $\Fva = \diag(\gamma_r,\gamma_\perp,\gamma_\perp)$. 

For small $h_0$, we expand $\tau_{rr} = f_{\text{ext}} (r-\rho+h)/h + O(h_0^2)$ and the mechanical balance along the radial direction gives:
$$
0 = [\grad\cdot\ts{\tau}]_r 
 = \frac{f_\text{ext}}{h} - \frac{4G}{r}(\alpha_\perp^2-1) 
$$
which results in an algebraic equation linking $\alpha_\perp$ and $\lambda_\perp$. As above, \eqref{eq-morpho-va} gives evolution equations for $(\gamma_r,\gamma_\perp)$ which allow to solve the dynamics (see \fig{fig-cyst}B) and find a nondegenerate steady state for $a^2\leq a_c^2 = \lfrac{8h_0 G}{(\rho_0 E_s)}$, 
$$
\alpha_\perp=\sqrt{1+\frac{a^2}{2}},\quad
\alpha_r=1,\quad
\lambda_\perp = \frac{1}{2}\left(1+\sqrt{1-\frac{a^2}{a_c^2}}\right),\quad
\lambda_r = 1/\lambda_\perp^2.
$$
For $a>a_c$, the shell collapses (\fig{fig-cyst}C).
Qualitatively, we obtain analogous phenomena as in the above case of the active beam, with a rich dynamical behaviour which can be solved in a convenient way thanks to the deformation gradient decomposition.

\section{Conclusions}

In this paper, we derive the specific shape and dependences on microscopic processes of
active terms that are present in linearisations of the active gel theory
\cite{Kruse+.2005.1} while ensuring that they are consistent with
thermodynamical requirements. Rather than start from the free energy of nematic liquid
crystals \cite{Kruse+.2004.1,Kruse+.2005.1}, where no stretching of the microstructure
is possible, we start from the entropic elasticity of Gaussian chains that model the actin
filaments between two crosslinks. This possibility of stretching the filaments is at
the origin of the specific shape of the objective derivative in the constitutive equation
\cite{Hinch-Harlen.2021.1}, whose nonlinearities are negligible only in the purely viscous 
limit. Thus, whenever the timescale of the process at play is comparable to the relaxation
time, the correct form of the constitutive equation is an upper-convected viscoelastic
liquid material law.
The model we derive here has a single relaxation time, however 
similar extensions as those of the Lodge network model \cite[p. 120]{Larson.1999.1} can be relevant for biopolymers
\cite{Broedersz+MacKintosh.2010.1}, and could allow to fit the fractional exponents observed 
in experiments \cite{Trepat+Fredberg.2008.1,Bonfanti+Kabla.2020.1}.
The simplest of these laws, when only one relaxation time is present---either due to the
material itself or because the timescale of the process is comparable but larger than
the largest relaxation time---is the upper-convected Maxwell equation.

A link is explicitely made between the microscopic scale
behaviour of molecular motors and the continuum scale active stress.
Compared to a previous such attempt \cite{Etienne+Asnacios.2015.1}, we are now
able to define conditions for which microscopic scale kinetics are
thermodynamically admissible. When motors are assumed to walk randomly along
actin filaments, we recapitulate the results from \cite{Etienne+Asnacios.2015.1},
reaching an anisotropic contractility which scales as the square of the length of
the typical step performed by a power-stroke of the myosin. We can also envision
a case where motors walk processively along polar filaments. The resulting shape of
the anisotropic contractility remains similar as in the above case, but with a linear
dependence in the step size and thus a higher efficiency.
We show that this active contribution appears as an offset of the isotropic rest configuration
of the microstructure to a new configuration with prestrained microstucture configuration.

Active biopolymer networks encompass both subcellular structures such as the
actomyosin and, at a larger scale, fibrous tissue \cite{Erlich+Wyatt.2022.1}. Contractility
of actomyosin is also felt in cellularised tissue like epithelia 
\cite{Khalilgharibi-Fouchard+.2016.1,Erlich+Wyatt.2022.1} which are often
simulated with similar continuum models \cite{Dicko+Etienne.2017.1,Wyatt+Charras.2020.1}.
For those it is not appropriate to use elastic chains as the microstructure,
however when considering the kinetics of cell deformation and neighbour exchanges,
models of the same shape as the present model are found, including the upper-convected
objective derivative \cite{Tlili+Saramito.2015.1,Ishihara+Sugimura.2017.1,Bandil-Vernerey.2023.1}.
A possible direction for these materials is to consider a non-Hookean elasticity of
the microstructure, e.g.\ with a finite extensibility approach \cite[p. 142]{Larson.1999.1}.

Viscoelastic liquids lead to mathematical problems notoriously difficult to solve
analytically or numerically, and the set of equations in \refsec{sec-summary} can
prove challenging to solve for complex geometries. In addition, biopolymer networks
such as actomyosin often form thin shell-like structures \cite{Erlich+Wyatt.2022.1},
which turns the problem into a moving domain partial differential system. 
We propose to use the formalism of multiplicative strain decomposition, often used 
for plasticity or morphoelastic descriptions of growth \cite{Lubarda.2004.1}
but which has
also proven useful e.g.\ to model anisotropic solid viscoelastic tissues
\cite{Ciambella+Nardinocchi.2024.1}.
We show that this allows to define a tractable resolution procedure for large deformations.

\section*{Acknowledgments}
J.E. is grateful to John Hinch, Claude Verdier and Atef Asnacios for their important
contributions to his approach of this topic.
The authors thank Alexander Erlich, Eric Bertin, Jonathan Fouchard, Catherine Quilliet 
and an anonymous reviewer for their helpful comments.

\appendix

\section{Asymptotic analysis of the dynamics of the unbound chains}
\label{app-unbound}

\JEadds{
We choose the characteristic time $\tau_b = \kb^{-1}$, and nondimensionalise with $\tsL = \overline{\tsL}/\tau_b$, and $\vc{r}=\beta^{-1}\overline{\vc{r}}$.
%\JE{Introduce a specific charac time, say $\dot\gamma$, for $\tsL$? Not so much better, since there's K too and prevents factoring out}
Then \eqref{eq-smolu-psiu} writes:
}
\JEadds{
\begin{align*}
    \ded{\psiu}{\overline{t}}
          + \gradrbar\cdot\left( \psiu \overline{\tsL}\cdot\overline{\vc{r}} 
            - \frac{\tau_b\kappa}{\zeta} \psiu\overline{\vc{r}} 
            - \frac{\tau_b k_BT\beta^2}{\zeta} \gradrbar\psiu \right)
      = \tau_b\ku\mathcal{K}_u(\psiu,\psib) %\psib - \overline{\kb}\psiu
\end{align*}
}
%\JE{CAn put back $\mathcal{K}_u$ instead}
\JEadds{
noting that $\kappa=k_BT\beta^2$, this can be rearranged as:
\begin{align*}
  \gradrbar\cdot(\psiu\overline{\vc{r}} 
  + \gradrbar\psiu)
  = \frac{\zeta}{\tau_b\kappa}\left(
    \ded{\psiu}{\overline{t}}
          + \gradrbar\cdot( \psiu \overline{\tsL}\cdot\overline{\vc{r}})
             -\tau_b\ku\mathcal{K}_u(\psiu,\psib) %\psib  + \overline{\kb}\psiu 
    \right)
\end{align*}
Here the nondimensional number $\lfrac{\zeta}{(\tau_b\kappa)}$ compares the
rate of binding $\kb = 1/\tau_b$ to the rate of equilibration of spring
and Brownian forces in unbound chains, $\lfrac{\kappa}{\zeta}$. We 
assume that this number is vanishingly small. We also assume that $\tau_b\ku\mathcal{K}_u$ and $\overline{\tsL}$ are of order 1 at most, we come back to these assumptions below. Then we can expand 
$\psiu(\vc{r},t) = \au(t)\psi_0(\vc{r}) + (\lfrac{\zeta}{(\tau_b\kappa)})\psiu'(\vc{r},t)$, where 
$\au = \intd{\Reals^3}{}{\,\psiu}{\pt{r}}$ and
the constant distribution $\psi_0(\vc{r})$ is such that $-\kappa \psi_0 \vc{r} - k_BT\gradr\psi_0 = \vc{0}$, thus solving the left-hand side,
$$
\psi_0 = (\kappa/(2\pi k_BT))^{3/2} \exp(-\kappa \pt{r}^2/(2k_BT)).
$$
The deviation $\psiu'$ solves:
\begin{align*}
  \gradrbar\cdot(\psiu'\overline{\vc{r}} 
  + \gradrbar\psiu')
  = \ded{\au}{\bar{t}}\psi_0 +
    \gradrbar\cdot( \psi_0 \overline{\tsL}\cdot\overline{\vc{r}}) 
              -\tau_b\ku\mathcal{K}_u\left(\psi_0+\frac{\zeta}{\tau_b\kappa}\psiu',\psib\right) %\psib + \overline{\kb}\psi_0 
    + \frac{\zeta}{\tau_b\kappa}
      %\left(
            \gradrbar\cdot( \psiu' \overline{\tsL}\cdot\overline{\vc{r}}).
      %\right).
\end{align*}
}
\JEadds{
Integrating over $\Reals^3$, noting that $\psiu'$ integrates to zero and using the divergence theorem,
$$
\ded{\au}{\bar{t}} =  \tau_b\ku\intd{\Reals^3}{}{\mathcal{K}_u(\psiu,\psib)}{\bar{\pt{r}}}.
$$
Since we have chosen $\mathcal{K}_u = \psib - (\kb/\ku)\psiu$, and $\intd{\Reals^3}{}{\,\psib}{\bar{\pt{r}}} = 1-\au$, we obtain
$$
\au(t) = \frac{\ku}{\kb+\ku} + \au^0 \exp(-(\ku+\kb)t)
$$
where $\au^0=\au(0)-\lfrac{\ku}{(\kb+\ku)}$.
}
\JEadds{
Thus, after a transient, $\au$ is a constant and $\psiu'$ 
is such that
\begin{align*}
  \gradrbar\cdot(\psiu'\overline{\vc{r}} 
  + \gradrbar\psiu')
  = \gradrbar\cdot( \psi_0 \overline{\tsL}\cdot\overline{\vc{r}}) 
             -\tau_b\ku\mathcal{K}_0(\psib)
    + O\left(\frac{\zeta}{\tau_b\kappa}\right)
\end{align*}
with $\mathcal{K}_0(\psib) = \psib - (\kb/\ku)\psi_0$ and is at most of order 1 if $\max\{\overline{\tsL},\ku/\kb\}$ is. 
Finally, this asymptotic development is valid for $\max\{ |\tsL|, \ku \} \lesssim \kb \ll \kappa/\zeta$. The transient is of duration at most $1/\kb$, thus shorter than the flow characteristic time $1/|\tsL|$.
}

\section{Diffusive molecular motors}
\label{app-motors}

In \refsec{sec-motors} we assume that molecular motors can sense a directionality in the
network and follow the oriented vector $\vc{r}$. This corresponds to the behaviour
of myosin minifilaments on actin networks, however this may not be the case of all contractile 
biopolymer networks. Starting again from \eqs{eq-onsager}, it is possible to construct another 
vectorial reaction coefficient based on the gradient of $\psib$, 
i.e.\ for any orientation tensor $\ts{q}$, specify $\Ja$ such that
it will yield a diffusion term in \eqref{eq-smolu}:
\begin{align*}
\lambda_{12}^{\text{diff}} &= -\frac{\ka\theta^{\text{diff}}}{\kappa} \ts{q}\cdot\gradr \log \psib,
&
\Ja^{\text{diff}} &= 
  	-\frac{\ka\theta^{\text{diff}}\Deltamu}{\kappa}
  \ts{q}\cdot\gradr \log \psib - \frac{1}{2}\ks\vc{r},
\\
&
&
\dxic^{\text{diff}} &= 2\ka\vc{r}\cdot\ts{q}\gradr\log\psib + \lambda_{22}^{\text{diff}} \Deltamu,
\end{align*}
%For simplicity we have taken $\ks=0$. 
where $\ka$ is a typical rate of motor progression and
$\theta^{\text{diff}}$ a numeric constant. As in the case of processive motors, we have taken $\lambda_{11} = \ks/(2\kappa)$.
In \cite{Etienne+Asnacios.2015.1}, we argue that
the resulting diffusion term in \eq{eq-smolu} must scale with the square of the size of
the steps $\ell_{\mathrm{m}}$ that the molecular motors perform on the strands, thus
$\theta^{\text{diff}}=\ell_{\mathrm{m}}^2\beta^2$. Additionally, we can identify that $\va = \ell_{\mathrm{m}}\ka$, where $\va$ was defined in \refsec{sec-motors} and found to be equal to $\ella(\ku+\ks)$. Evaluating $\lambda_{22}^{\text{diff}}$ is more difficult than $\lambda_{22}$ in the processive case, however taking $\psib$ to be a small deviation from $\psi_0$, we find $\lambda_{22}^{\text{diff}}>3((\ku+\ks)\ella\ell_{\mathrm{m}}\beta^2)^2/(16 \pi \ks k_BT)$.

If we choose $\ts{q}=\ts{I}$, since $\langle \vc{r}\gradr\log\psib \rangleb = -\frac{\kb\au}{\ku}\ts{I}$, 
the %average rate of advancement of the reaction is 
%$\langle\dxic^{\text{diff}}\rangleb=\lambda_{22}\Deltamu-2\va^{\text{diff}}\beta$. The 
contractility tensor %and the active stress %in this case are:
is:
\begin{align*}
\tAa &= \frac{2\ella\ell_{\mathrm{m}}\beta^2\kb\au\Deltamu}{\ku\kappa} \ts{I}.
%&
%\tsiga &=...
\end{align*}
However the lack of alignment of the velocity $\Ja$ with the microstructure is difficult to interpret.
Following \cite{Etienne+Asnacios.2015.1}, 
it is also possible to take $\ts{q}=\ts{Q}$, which means that the diffusive-like behaviour of
the molecular motors takes place along the direction of the strands $\vc{r}/|\vc{r}|$.
Thanks to the property $\langle \vc{r}\ts{Q}\gradr\log\psib \rangleb = -\langle\ts{Q}\rangleb$,
as above, and to the fact that $\tr\ts{Q}=1$, we have 
%$\langle\dxic^{\text{diff}}\rangleb=2\ka^{\text{diff}}+\lambda_{22}$ and is thus
%independent of the mechanical stress.
%The contractility tensor and the active stress are then:
\begin{align*}
\tAa &= \frac{2\ella\ell_{\mathrm{m}}\beta^2\Deltamu}{\kappa} \langle\ts{Q}\rangleb.
%&
%\tsiga &={2\nu\ell_{\mathrm{m}}^2\beta^2\Deltamu} \frac{\ka}{\ku} \langle\ts{Q}\rangleb.
\end{align*}

Thus we find that the ``diffusive'' and ``processive'' types of reactive flux $\Ja$ lead to
expressions of the active stress which
are highly similar. 
%\JE{calculate efficiencies? seems not so interesting since I have no idea whether $\ell_m\beta$ should be $>1$}
%differ only by the factor %$\ell_{\mathrm{m}}^2\beta/\ell_{\mathrm{a}}$.
%The distance $\ell_{\mathrm{a}}$ over which a myosin motor is %processive can be expected to be much greater than
%the Kuhn length $\beta^{-1}$ and motor steps $\ell_{\mathrm{m}}^2$, %thus the processive behaviour is much more efficient.

\section{Lower bound on the trace of the microstructure tensor}
\label{app-trace}

\JEadds{
We show here that for appropriate initial conditions, and in particular $\ts{A}(t=0)=\ts{A}_0$,
the trace of $\ts{A}$ remains larger than the one of $\ts{A}_0$. We use the fact that the material is incompressible ($\gradx\cdot\vc{v}=0$) and impose that
$\tAa$ is positive semi-definite, hence corresponds to a contractile active term.
As stated in the main text, we essentially follow the proof of Lemma 2.1 by \citetext{Hu-Lelievre.2007.1}
but take into account the additional active term.
}

\JEadds{
The modified lemma claims: assume that $\det\ts{B}(t=0) \geq 1$ and 
$\tau\ucm{B} = - \ts{B} + \ts{I} +\ts{B}_a(t)$ with $\ts{B}_a(t)$ a symmetric semi-definite tensor. Then, $\forall t>0$, $\det\ts{B}(t) \geq 1$.
We apply this lemma with $\tau=\ku^{-1}$, $\ts{B} = \beta^2 a_0^{-1} \ts{A}$ and $\ts{B}_a = \beta^2 a_0^{-1} \tAa$, which matches the conditions of \eqref{eq-relaxtive}. Note that 
both are indeed symmetric semi-definite by construction, since $\ts{A} = \langle \vc{r}\vc{r} \rangleb$ and $\tAa = \frac{2\ell_a \beta\Deltamu}{\kappa}\langle\frac{\vc{r}\vc{r}}{\pt{r}^2}\rangleb$. 
The condition $\det(\ts{B}(t=0)) \geq 1$ is equivalent to $\det(\ts{A}(t=0)) \geq \det(\ts{A}_0)$.
}

\JEadds{
Following \citetext{Hu-Lelievre.2007.1}, we observe that using Jacobi's formula, and for incompressible flow ($\tr\tsL=0$),
\begin{align*}
    \tau\dedt{\ln(\det(\ts{B}))} + \tau\vc{v}\cdot\gradx\ln(\det(\ts{B}))
    &= \underbrace{\tau\tr\left(\tsL+\tsL\trsp\right)}_{=0}
      +\tr\left( - \ts{I} + \ts{B}^{-1} + \ts{B}^{-1}\ts{B}_a \right)
\end{align*}
We use the inequality of arithmetic and geometric means applied to the non-negative eigenvalues of $\ts{B}^{-1}$ and $\ts{B}^{-1}\ts{B}_a$,
$$\frac{1}{3}\tr \ts{M} \geq  (\det\ts{M})^{1/3},$$
and, introducing the derivative $\ddt{}$ along flow characteristics, obtain
\begin{align*}
    \frac{\tau}{3}\ddt{\ln(\det(\ts{B}))}
    &\geq - 1 + \det(\ts{B})^{-1/3}\left(1 + \det(\ts{B}_a)^{1/3}\right)
    \\
    &\geq -1 + \det(\ts{B})^{-1/3}
\end{align*}
Setting $z=1-\det(\ts{B})^{1/3}$, we find $z(0)\leq 0$ and $\tau\ddt{z} \leq -z$, thus $z(t) \leq z(0) e^{-t/\tau}$ and $\det(\ts{B}(t)) \geq 1$.
}

\JEadds{
Using the above upper bound of the determinant, 
$$
\tr \ts{A}_0 \leq 3a_0\beta^{-2} \det(\beta^2 a_0^{-1} \ts{A})^{1/3} \leq \tr \ts{A}.
$$
}

\bibliographystyle{apalike}      
\bibliography{main}   
%\printbibliography%

\end{document}